\documentclass[pre,twocolumn,superscriptaddress,floatfix,aps]{revtex4-2}
\usepackage{graphicx} 
\usepackage{amsmath,amssymb,bm}
\usepackage{graphicx}
\usepackage{physics}
\usepackage[T1]{fontenc}
\usepackage{comment}
\usepackage{float} 
\usepackage{mathtools}
\usepackage{tabularx}
\usepackage{multirow}
\usepackage{xcolor}
\usepackage[normalem]{ulem}
\usepackage{hyperref}
\hypersetup{colorlinks,%
linkcolor=blue,%
citecolor=blue,%
urlcolor=blue}
\usepackage{color}
\usepackage{tikz}
\usetikzlibrary{decorations.markings, arrows.meta, calc}

\begin{document}

\title{A robust method \textcolor{black}{to identify} chimera states}

\author{S. Nirmala Jenifer}
\email{nirmala-jenifer.selvaraj@unamur.be}
\affiliation{Department of Physics, Bharathidasan University, Tiruchirappalli 620 024, Tamil Nadu, India}
\affiliation{Department of Mathematics and naXys, Namur Institute for Complex Systems, University of Namur, Namur, Belgium}

\author{Riccardo Muolo}
\affiliation{RIKEN Center for Interdisciplinary Theoretical and Mathematical Sciences (iTHEMS), Saitama, Japan}
\affiliation{Department of Systems and Control Engineering, Institute of Science Tokyo (former Tokyo Tech), Tokyo, Japan}

\author{Paulsamy Muruganandam}
\affiliation{Department of Physics, Bharathidasan University, Tiruchirappalli 620 024, Tamil Nadu, India}
\affiliation{Department of Medical Physics, Bharathidasan University, Tiruchirappalli 620 024, Tamil Nadu, India}

\author{Timoteo Carletti}
\email{timoteo.carletti@unamur.be}
\affiliation{Department of Mathematics and naXys, Namur Institute for Complex Systems, University of Namur, Namur, Belgium}

\begin{abstract}
Chimera states are one of the most intriguing phenomena in nonlinear dynamics, characterized by the coexistence of coherent and incoherent behavior in systems of coupled identical oscillators. Despite extensive studies and numerous observations in different settings, the development of reliable and systematic methods to classify chimera states and distinguish them from other dynamical patterns remains a challenging task. Existing approaches are often limited in scope and lack robustness. In this work, we propose a method based on Fourier analysis combined with statistical classification to {\color{black}identify} chimera behavior. The method is applied to a system of topological signals coupled via the Dirac operator, where it successfully captures the rich dynamical regimes exhibited by the model. We demonstrate that the proposed approach is robust with respect to variations in network topology and system parameters. Beyond the specific model considered, the framework provides a general and automated tool for distinguishing different dynamical regimes in complex systems.
\end{abstract}
\maketitle

\section{introduction}

Complex networks provide a natural framework for describing large collections of interacting dynamical units whose collective behavior cannot be inferred from individual components alone~\cite{Latorabook}. Such systems arise in diverse contexts, by including neural populations, chemical oscillators, power-grid dynamics, and ecological interactions, where the interplay between intrinsic dynamics and network structure gives rise to rich spatiotemporal patterns~\cite{boccaletti2006complex, arenas2008synchronization, Laing2009a, Sawicki2023}. Among these emergent behaviors, chimera states, characterized by the coexistence of coherent and incoherent dynamical regions within a population of identical oscillators, have attracted sustained interest due to their counterintuitive nature and potential relevance to real-world phenomena.

Chimera states were first reported by Kaneko for coupled maps~\cite{kaneko1984period,kaneko1}, and later observed for global~\cite{hakim1992dynamics, nakagawa1993collective, chabanol1997collective}, i.e., all-to-all, and nonlocal~\cite{kuramoto1995scaling, kuramoto1996origin, kuramoto1997power, kuramoto1998multiaffine, kuramoto2000multi}, i.e., first neighbors, couplings. However, such patterns became popular after the studies on phase oscillators by Kuramoto and Battogtokh~\cite{kuramoto_batt}, and by Abrams and Strogatz~\cite{Abrams2004}, who coined its actual name. Since then, they have been observed in a wide range of models and experimental systems, for instance, in Josephson junction arrays~\cite{cerdeira_prl}, electronic circuits~\cite{vale_chimeras, gambuzza2020experimental}, lasers~\cite{hagerstrom2012experimental}, mechanical oscillators~\cite{martens2013chimera}, and nano-electromechanical systems~\cite{matheny2019exotic}. They have sparked also interest beyond the nonlinear science community, since chimera-like patterns have been suggested as models for unihemispheric sleep observed in certain animals~\cite{chimera_neuro, majhi2019chimera, rattenborg2000behavioral}. Strictly dynamically speaking, their importance lies in the fact that they provide a concrete example of partial synchronization, where identical units subject to identical coupling self-organize into distinct dynamical groups~\cite{Panaggio2015, zakharova2020chimera, Parastesh2021}, and, nowadays, several kinds of patterns have been observed, such as amplitude chimeras~\cite{zakharova2014chimera}, amplitude-mediated chimeras~\cite{sethia2013amplitude}, and phase chimeras~\cite{zajdela2025phase}, to name a few.

Despite their conceptual importance, reliably identifying and characterizing chimera states remains a nontrivial task. \textcolor{black}{Many results are based on the observation of an ``arc-like'' distribution of mean-phase velocities, where, roughly, the flat part corresponds to coherent and the arc to the incoherent region~\cite{omelchenko2019control}. This behavior has been considered to be a signature for identifying chimeras. However, when quantifying the variation of the model parameters to observe the emergence of chimera states, such a qualitative method may not be effective. A variety of measures have been thus proposed to quantify/identify chimeras and their dependence on relevant parameters, e.g., the complex
order parameter~\cite{martens2013chimera}, the standard deviation of mean phase velocity~\cite{omelchenko2016tweezers}, the distance between the center of
mass of each oscillator and the origin~\cite{zakharova2014chimera}, strength of Incoherence~\cite{gopal2014} and total normalized variations of phases \cite{muolo2024phase}.} 

Among those methods, the \emph{Strength of Incoherence} $(\mathrm{SI})$~\cite{gopal2014} has been largely used; it ultimately captures the system behavior in a single scalar: coherent states are associated to $\mathrm{SI}=0$, chimera states to $0<\mathrm{SI}<1$, and incoherent ones to $\mathrm{SI}=1$. 
\textcolor{black}{Let us emphasize that this method relies on the (arbitrary) choice of some thresholds, which impact  the value of $\mathrm{SI}$ and, thus, the resulting identification of chimera or lack thereof (see Appendix~\ref{sec:compS})}. Moreover, $\mathrm{SI}$ does not convey information about  the spatial dependence of frequency, amplitude, and phase, which is the very first signature of chimera states. Generally, the existing approaches often depend sensitively on parameter choices, system size, thresholds or {\color{black} the type of existing chimera states in the system; in general, it is not possible to use the same set of method parameters to compute the $\mathrm{SI}$ and identify different kinds of chimera states, i.e., phase chimeras and amplitude chimeras}, and may lead to ambiguous interpretations when applied to weak, transient, or spatially irregular patterns. In practice, the distinction between chimera and non-chimera regimes is not always sharply defined, and intermediate behaviors can be difficult to categorize in a consistent and robust way. Because the problem does not have a definite solution yet, the development of signal-based approaches that can capture gradual variations in local correlations without relying on ad hoc criteria, opens new interesting research avenues.


In this work, we consider a system of topological signals defined on a $1$-simplicial complex, i.e., a network, where dynamical variables are associated not only with nodes but also with links~\cite{bianconi2021higher}. The model fits, thus, in the recent framework of higher-order networks~\cite{battiston2020networks, majhi2022dynamics, bick2023higher, boccaletti2023structure, battiston2026collective}, where interactions are not limited to pairwise connections but can involve more complex relations encoded by hypergraphs, and simplicial or cell complexes~\cite{natphys, millan2025topology}. More specifically, we consider the dynamics of topological signals coupled via the discrete Dirac operator~\cite{bianconi2021topological}, such that the nodes behave as coupled oscillators, despite having no intrinsic oscillatory dynamics. Such setting has been shown to yield rich synchronization~\cite{carletti2025global, muolo2026synchronization} and pattern formation~\cite{giambagli2022diffusion, muolo2024three, muolo2024turing} dynamics. The adapted FitzHugh-Nagumo model~\cite{FitzHugh1961, Nagumo} considered here, with dynamics on both nodes and links, exhibits a rich range of behaviors, by including chimera, coherent and incoherent regimes, making it a suitable benchmark for classification tasks. Note that, while chimera states have been found on hypergraphs~\cite{zhang2021unified, kundu2022higher, ghosh_chimera2, bick_nonlocal1, zhang2024deeper, mau2024phase, muolo2025pinning, djeudjo2025chimera}, such behavior had not been observed in systems of coupled topological signals. 

Motivated by this research question, we develop a Fourier transform-based framework to analyze the time series, i.e., the time evolution of node signals; by starting from simulated data, we extract instantaneous amplitude, phase, and frequency for the time series generated from the dynamics of each node by using a refined windowed Fourier transform. We then quantify their local variations across the network by computing the total normalized variations~\cite{muolo2024phase}, a measure of the (spatial) smoothness of the signal under study. In this way, we associate to each time series three quantities, the total normalized variation of amplitude, phase and frequency, that form the basis of a classification scheme capable of distinguishing between synchronized, chimera, and irregular regimes. More precisely, we leverage on the use of hierarchical clustering and dendrograms to cluster the data and thus automatically determine the class and the associated dynamical behavior. The proposed approach provides a robust way to identify different dynamical patterns by using directly measurable signal features. {\color{black} Moreover,} we have shown that it does not rely on a given threshold as, e.g., for the SI, and it is applicable to a broad class of networked dynamical systems, beyond the ones hereby proposed.

The remainder of the manuscript is organized as follows. Sec.~\ref{sec:model} introduces the theoretical model, the governing equations of the system, and shows the existence of the homogeneous solutions along with their stability analysis. Sec.~\ref{sec:results} presents the main results, the Fourier method and the classification. We conclude Sec.~\ref{sec:results} by exploring the effects of system parameters on the observed dynamical regimes. Finally, in Sec.~\ref{sec:concl} we resume our main findings and discuss possible future research directions.

\section{The model}   
\label{sec:model}


We consider a ring composed by $\color{black}{N_0}$ nonlocally coupled identical FitzHugh-Nagumo oscillators (FHN)~\cite{FitzHugh1961, Nagumo, Rinzel}, each one described  by two dynamical variables, the membrane potential, $u_i$, and the recovery variable, $v_j$. 
This model has been proposed to study the dynamics of neurons, with neurons interacting with synaptic variables and vice-versa~\cite{ermentrout2010mathematical}. In this work, we assume the membrane potential to be anchored to each node, while the recovery variable is associated to each link; let us observe that a similar model has been proposed in~\cite{muolo2026synchronization}. This assumption naturally sets our model in the framework of topological signals defined on simplicial complexes, where, i.e., a node is a $0$-simplex and a link a $1$-simplex. Moreover, simplices of different dimensions are coupled through the Dirac operator~\cite{bianconi2021topological, giambagli2022diffusion}. Our system naturally includes simplices up to dimension $1$, hence involving only pairwise interactions. It can, however, be extended to higher-dimensional simplicial complexes to represent higher-order, non-pairwise interactions such as three-body or four-body interactions~\cite{battiston2020networks, majhi2022dynamics}, as well as on cell-complexes~\cite{schaub2021signal, carletti2023global}. 


Based on the above, the equations ruling the system evolution, comprising node and link dynamics coupled by the Dirac operator~\cite{bianconi2021topological,giambagli2022diffusion}, are given by
\begin{align}
\textcolor{black}{\epsilon}\frac{d{u}_i}{dt} &= u_i - \frac{u_i^3}{3} - (\mathbf{B}_1 v)_i\, , \label{eq:eq1a}\\ 
\frac{d{v}_j}{dt} &= b + c v_j + (\mathbf{B}_1^\top u)_j,
\label{eq:eq1b}
\end{align}
where $\mathbf{B}_1$ is the incidence matrix~\cite{Lim2020} of size $\color{black}{N_0} \times \color{black}{N_1}$ for a system with $\color{black}{N_0}$ nodes and $\color{black}{N_1}$ links. 

Because each node is connected to $P$ neighbors on either side, the resulting node degree is $k_i = 2P$, for all $i=1,\dots,\color{black}{N_0}$. Moreover, we assume the orientation of the network~\footnote{\textcolor{black}{Let us recall that to define the incidence matrix, it is mandatory to orient the links; let us also stress that the latter is not related to any link directionality, and indeed the network is symmetric.}} to be clock-wise: namely a node, $i_1$, is the tail of the oriented link $[i_1,i_2]$ for any $i_2$ ``being ahead on the clock face'' $i_1$ such that $|i_1-i_2|\leq P$, $i_2$ is thus the head of the link~\footnote{Formally, this can be restated as follows. Assume the nodes labels to be integer numbers in $\{1,\dots,N_0\}$ and assume for sake of simplicity $N_0$ to be even; let $I=\{1,\dots,N_0/2\}$ and $J=\{N_0/2+1,\dots,N_0\}$, if $i_1$ and $i_2$ belong to different sets, i.e., $i_1\in J$ and $i_2\in I$ then ``$i_2$ is ahead of $i_1$'' means $i_2>i_1-N_0$. In the remaining case, the order relation is the usual one, i.e., $i_2>i_1$}, the incidence matrix $\mathbf{B}_1$ can thus be computed as follows: given an oriented link \textcolor{black}{$j=[i_1,i_2]$},
\textcolor{black}{
\begin{align}
\label{eq:B1initial}
\mathbf{B}_1[i,j] =
\begin{cases}
\;\;1, &\quad i = i_2, \\
-1, & \quad i = i_1, \\
\;\;0, & \text{otherwise}.
\end{cases}
\end{align}}
Finally, let us observe that the number of incoming and outgoing links of each node is the same; we refer to this configuration as \emph{orientation~1} (see left panel of Fig.~\ref{fig:orient} for an example with $\textcolor{black}{{N_0}} = 8$ and $P = 3$, {\color{black} from which one can clearly observe the presence of a rotation invariance in the structure of the network}).
\begin{figure*}[!ht]
\centering
\begin{tikzpicture}[scale=0.75, transform shape]
\def\radius{3cm}

\tikzset{
 mnode/.style={
 circle,
 draw=black,
 thick,
 minimum size=0.75cm,
 fill=teal!80, 
 text=white,
 font=\large\bfseries ,
 inner sep=0pt,
 outer sep=0.25pt
 },
 midarrow/.style={
 postaction={decorate},
 decoration={
      markings, mark=at position 0.6 with {\arrow{Stealth[length=3mm, width=2mm]}}
   }
 }
}

\begin{scope}[xshift=0cm]
 \foreach \i in {1,...,8} {
    \node[mnode] (n\i) at ({90 - (\i-1)*360/8}:\radius) {\i};
 }

 \foreach \i in {1,...,8} {
 \foreach \k in {1, 2, 3} {
 \pgfmathtruncatemacro{\j}{mod(\i + \k - 1, 8) + 1}
 
 \draw[thick, postaction={decorate, decoration={markings, mark=at position 0.50 with {\arrow{Stealth}}}}] (n\i) -- (n\j);
 }
 }
\end{scope}

\begin{scope}[xshift=8.0cm]
 \foreach \i in {1,...,8} {
 \node[mnode] (n\i) at ({90 - (\i-1)*360/8}:\radius) {\i};
 }

 \foreach \i in {1,...,8} {
 \foreach \k in {1, 2, 3} {
 \pgfmathtruncatemacro{\j}{mod(\i + \k - 1, 8) + 1}
 \pgfmathparse{(\i==1 && \j==2) || (\i==8 && \j==1)}
 \ifnum\pgfmathresult>0
 \draw[thick, red, postaction={decorate, decoration={markings, mark=at position 0.50 with {\arrow{Stealth[reversed]}}}}] (n\i) -- (n\j);
 \else
 \draw[thick, postaction={decorate, decoration={markings, mark=at position 0.50 with {\arrow{Stealth}}}}] (n\i) -- (n\j);
 \fi
 }
 }
\end{scope}

\begin{scope}[xshift=16.0cm]
 \foreach \i in {1,...,8} {
 \node[mnode] (n\i) at ({90 - (\i-1)*360/8}:\radius) {\i};
 }

 \foreach \i in {1,...,8} {
 \foreach \k in {1, 2, 3} {
 \pgfmathtruncatemacro{\j}{mod(\i + \k - 1, 8) + 1}
 \pgfmathparse{\i==1 || \j==1)}
 \ifnum\pgfmathresult>0
 \draw[thick, red, postaction={decorate, decoration={markings, mark=at position 0.50 with {\arrow{Stealth[reversed]}}}}] (n\i) -- (n\j);
 \else
 \draw[thick, postaction={decorate, decoration={markings, mark=at position 0.50 with {\arrow{Stealth}}}}] (n\i) -- (n\j);
 \fi
 }
 }
\end{scope}

\end{tikzpicture}
\caption{\textbf{Schematic illustration of orientation $1$, for the case $\textcolor{blue}{N_0}=8$ and $P=3$}. In the left panel, we report the original ring, where each one of the eight nodes is connected to $P = 3$ neighbors on either side. The middle panel refers to the case where $Q=1$ links incident to node $1$, have been reoriented (drawn in red). In the right panel, we show the case $Q=P=3$, where again we fixed node $1$ as reference (red links are the reoriented ones).}
\label{fig:orient}
\end{figure*}

The parameter $b$ in Eq.~\eqref{eq:eq1b} determines whether nodes exhibit oscillatory, $|b| < 1$, or excitable behavior, $|b| > 1$~\cite{Ramlow_2019}. Because the matrix $\mathbf{B}_1$ satisfies $\mathbf{B}_1(1,\dots,1)^\top=0$, one can prove the existence of three homogeneous equilibria $(u_i,v_j)=(\sqrt{3}, -b/c)$, $(u_i,v_j)=(-\sqrt{3}, -b/c)$ and $(u_i,v_j)=(0,-b/c)$, the first two are stable while the latter is unstable. {\color{black}Let us observe that the presence of stable equilibria with a large basin of attraction reduces the possibility of observing other dynamical behaviors, among which are chimera states. Moreover,} when $b = 0$, the system is invariant under the transformations $(u,v) \rightarrow (-u,-v)$ and, thus, the dynamics will not depend on the chosen orientation. However, this invariance is broken when $b \ne 0$. For the following analysis, we fix $b = 0.5$ and $\textcolor{black}{\epsilon} = 0.05$. The condition $c < 0$ ensures that the recovery variables decay when we silence the interactions of the latter with neurons; we thus set $c=-0.01$. 

{\color{black} In the following, we will study  ring structures with different orientations with respect to the one given by~\eqref{eq:B1initial}, and determine the possible impact of the number of reoriented links on the emergence of chimera states. As a preliminary step, let us study the stability of the above equilibria and the size of their basin of attraction.} 

Let us thus rewrite Eqs.~\eqref{eq:eq1a} and~\eqref{eq:eq1b} as
\begin{align}
\label{eq:eq1aprime}
\frac{d{u}_i}{dt} & = {\textcolor{black}{\epsilon'}}(u_i - \frac{u_i^3}{3}) - \textcolor{black}{\epsilon'}\sum_lB_{1}[i,l]v_l
 \\ 
\frac{dv_j}{dt} & = \sum_iB_{1}[i,j]u_i + b + cv_j 
\label{eq:eq1bprime}
\end{align}
where $\textcolor{black}{\epsilon'} = 1/\textcolor{black}{\epsilon}$. To determine the stability of the homogeneous solution $(u_i,v_j)=(u^*,v^*)$, where $(u^*,v^*)=(\pm\sqrt{3},-b/c)$ or $(u^*,v^*)=(0,-b/c)$, {\color{black}we perform a linear stability analysis (see Appendix~\ref{sec:stabhomeq}) allowing us to prove that the fixed point $(u^*,v^*)=(0,-b/c)$ is always unstable, whereas the fixed points $(u^*,v^*)=(\pm\sqrt{3},-b/c)$ are always stable, and this holds true for any $P$.}

\begin{figure}[!ht]
\centering
\includegraphics[width=0.99\linewidth]{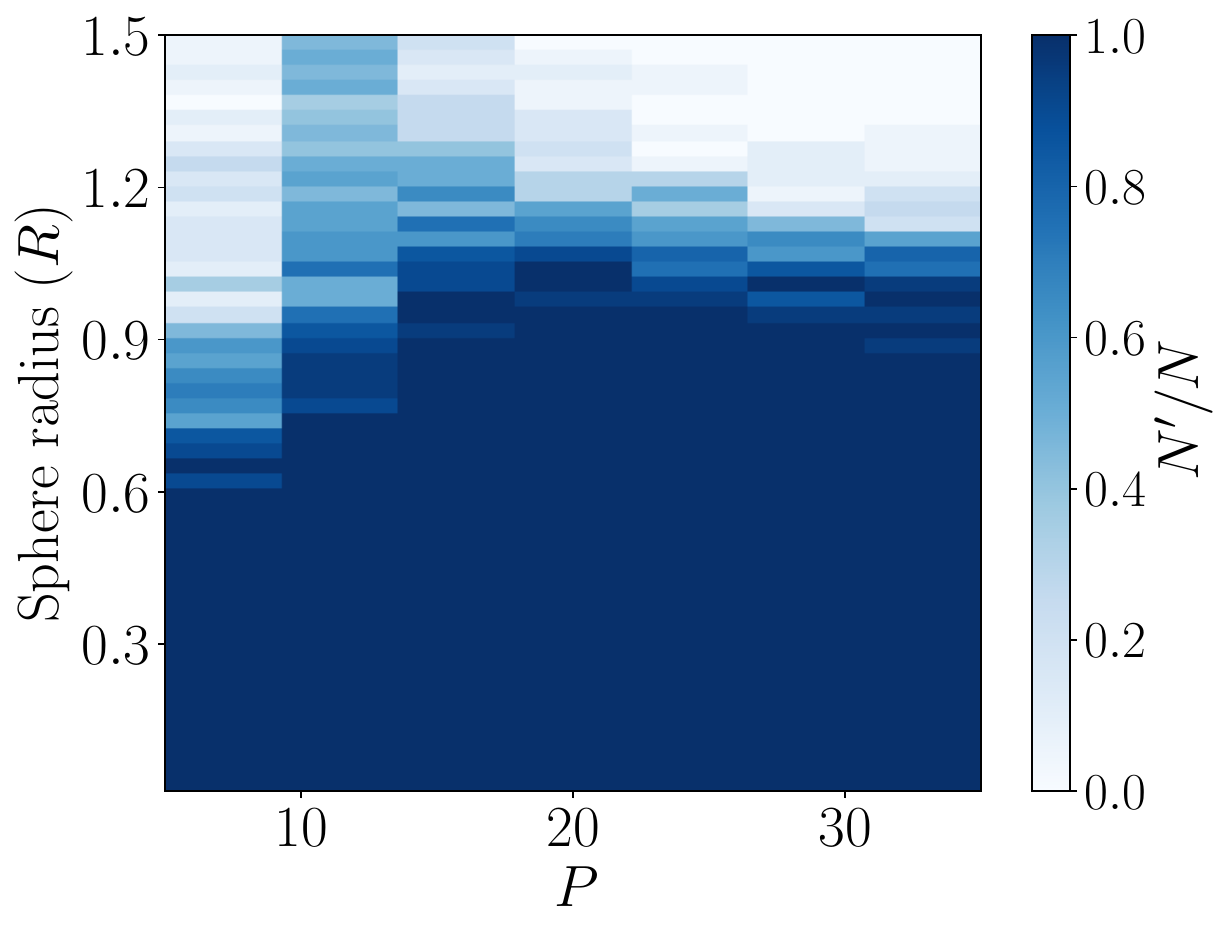}
\caption{\textbf{Stability domain of the homogeneous solution $u_i=\sqrt{3}$, $v_j=-b/c$ for the FHN model defined on a ring of $\textcolor{black}{N_0} = 100$ nodes with a variable number of links controlled by the parameter $P$}. The fraction $f = N'/N$ of initial conditions, starting within a ball of radius $R$ and remaining in the same ball after a sufficiently long time (i.e., staying ``close'' to the starting point), is shown as a function of $P$. A nonmonotonic behavior is observed: for small $P$, nearly $100\%$ of the orbits remain within a ball of radius $R \sim 0.6$. For larger $P$, the stability domain expands, with $f = 1$ up to $R \sim 1$ at $P = 20$. For even larger values of $P$, the stability domain slowly decreases.}
\label{fig:stability}
\end{figure}

To determine the {\color{black}size of the} domain of stability of the homogeneous solution, $(u_i,v_j)=(\sqrt{3},-b/c)$, we performed numerical simulations to determine the fraction, $f$, of initial conditions starting inside a ball of radius $R$, centered at the equilibrium, whose orbits remain in the same ball after a given interval of time. We are in particular interested in the dependence of $f$ on $P$ for a fixed network size. More precisely we fixed $\textcolor{black}{N_0}=100$ and for $P$ ranging in $[1,30]$, we considered $50$ values of $R$ in the interval $[0.01,1.5]$. For every radius, the simulation was repeated $N=20$ times with different initial conditions. Let $N'$ denote the number of realizations for which the solution remains within the prescribed radius $R$, we then compute the ratio $f=N'/N$. When $N'/N=1$, the equilibrium is stable for all realizations and it is therefore classified as stable. As the radius increases, $N'$ decreases, indicating the onset of instability (see Fig.~\ref{fig:stability}).

The existence of the homogeneous solution, where all $u_i$ converge to $\sqrt{3}$ and $v_j$ to $-b/c$ and its relatively large attraction basin, impedes the emergence of chimera states as well as other spatially heterogeneous solutions. For this reason, we decided to modify the links orientation and determine conditions to {\color{black}induce} the emergence of those states. Thus, by starting from the oriented ring with a given number of nearest neighbors for each node, $2P$, evenly distributed on the left and on the right, we inverted the orientation of $Q\in\{1,\dots,P\}$ links (see middle panel of Fig.~\ref{fig:orient} for the case $Q=1$ and the right panel where we show the case $Q=3$, in both cases $P=3$). Because of the invariance by rotation, all the nodes are equivalent, before to reorient the links, and thus we decided to focus on node $1$. Let us observe that now the homogeneous state is no longer a solution: indeed, the underlying structure does not exhibit anymore a rotation invariance.

Before to proceed with the presentation of the numerical results, let us observe that we can define a different orientation, where again the homogeneous state cannot be a solution. The latter structure is defined by using the incidence matrix
\begin{align}
\textcolor{black}{\hat{\mathbf{B}}_1}[\textcolor{black}{i_1},j] =
\begin{cases}
\;\;1, & \textcolor{black}{i_1} > \textcolor{black}{i_2}, \\
-1, & \textcolor{black}{i_1} < \textcolor{black}{i_2}, \\
\;\;0, & \text{otherwise}\, ,
\end{cases}
\end{align}
being $j=[\textcolor{black}{i_1},\textcolor{black}{i_2}]$ the oriented link according to the node index and $\textcolor{black}{\hat{\mathbf{B}}_1}[i_2,j] = -\textcolor{black}{\hat{\mathbf{B}}_1}[i_1,j] $. We refer to this case as orientation $2$. \textcolor{black}{In simple words, for a given link, {\color{black}$j=[i_1,i_2]$}, if node index $i_1$ is smaller than $i_2$, the (orientation of the) link will point from $i_1$ to $i_2$, and vice versa.} The interested reader can find more details about this orientation in Appendix~\ref{sec:orient2} and an example of a ring with $\textcolor{black}{N_0}=8$ nodes and $P=3$ is shown in Fig.~\ref{fig:orient:3}. By anticipating on the following, we will show that this second structure will be more robust with respect to the emergence of chimera state, once we randomly reorient {\color{black}a fixed number of} links (see Appendix~\ref{sec:orient2}).


\section{Method and results}   
\label{sec:results}

The aim of this section is to introduce and discuss the method used to classify the dynamical behaviors exhibited by the FHN system and the numerical results about the possible dynamical states exhibited by system~\eqref{eq:eq1a}-~\eqref{eq:eq1b} defined on the ring with orientation $1$. We first present our analysis in the case where we reorient the maximum allowed number of links, i.e., $Q=P$, still focusing on node $1$, and we let $P$ to vary from $1$ to $\textcolor{black}{N_0}/2-1$, being $\textcolor{black}{N_0}$ an even integer, {\color{black}or $(N_0-1)/2$ if $N_0$ is odd}. Then we will show the remaining case where also $Q$ varies in $\{0,\dots,P\}$. This analysis will be presented for a fixed set of model parameters, $\textcolor{black}{\epsilon}$, $b$ and $c$, their impact will be shortly studied next.

\begin{figure*}[!ht]
 \centering
  \includegraphics[width=1.005\linewidth]{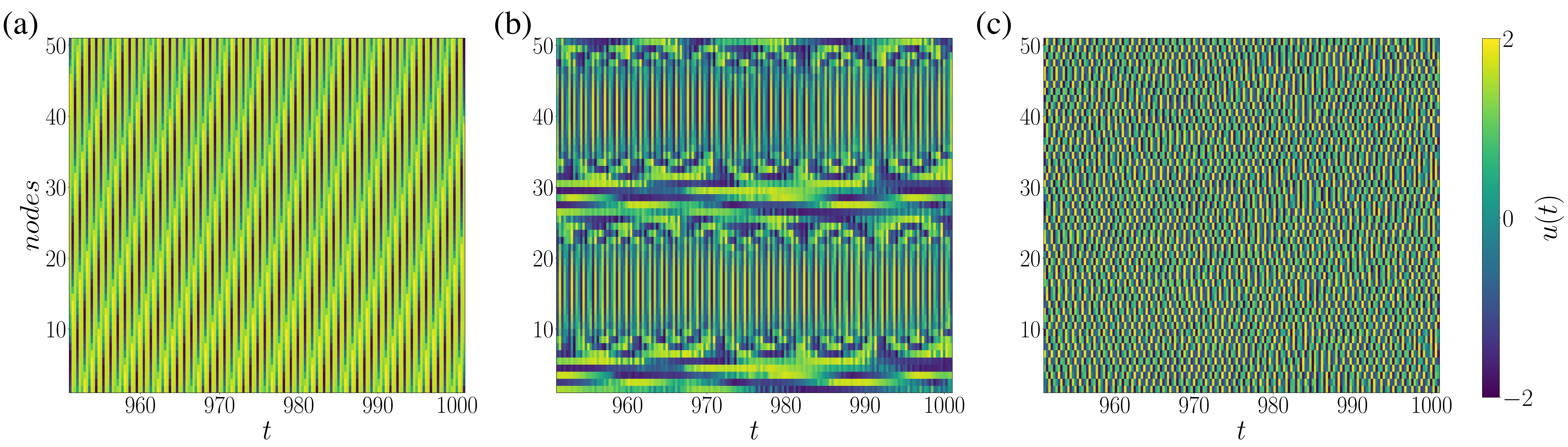}
  \caption{\textbf{Typical dynamical behaviors}. Numerical simulations of model~\eqref{eq:eq1a} and~\eqref{eq:eq1b} in orientation $1$, illustrating three distinct dynamical regimes: ordered behavior (panel (a), $Q=P=9$), a chimera state (panel (b), $Q=P=12$), and a disordered state (panel (c), $Q=P=21$). The remaining parameters are $n=50$, $\textcolor{black}{\epsilon} = 0.05$, $b = 0.5$, and $c = -0.01$. Simulations are performed using the Tsit5 solver over the time interval $[0,1000]$. For clarity, only the evolution of $u_i(t)$ over the final time window $t\in[950,1000]$ is shown.}
 \label{fig:9dyn}
\end{figure*}

\begin{figure*}[!ht]
 \centering
\includegraphics[width=0.97\linewidth]{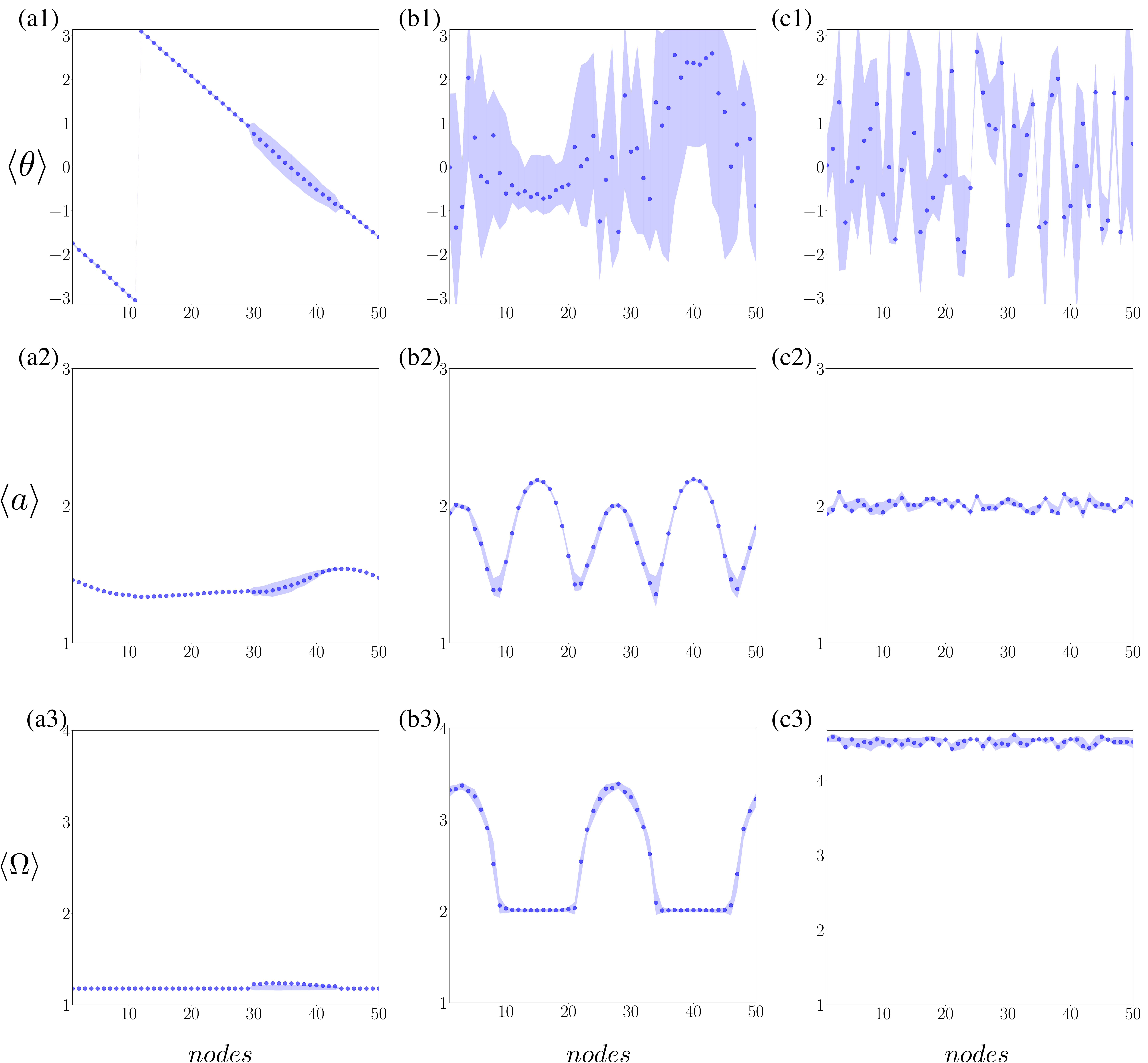}
 \caption{\textbf{Fourier indicators}. Average phases $\langle \theta\rangle$ (top row), average amplitudes $\langle a\rangle$ (middle row), and average frequencies $\langle \Omega\rangle$ (bottom row) obtained from the time series shown in Fig.~\ref{fig:9dyn} for $P=9$, $P=12$, and $P=21$. Column (a) shows ordered behavior, with a smooth dependence of all three quantities on the node index. Column (b) displays a chimera state, characterized by the typical distribution of frequencies; here, the average amplitudes remain smooth, while the phases exhibit less regularity. Column (c) shows a disordered state, marked by small but irregular variations in both amplitude and frequency, and highly irregular phase behavior. The corresponding variation values $(V_\theta, V_a, V_\Omega)$ are $(0.0683, 0.082, 0.0022)$ for the ordered state in column (a), $(0.2301, 0.1110, 0.1087)$ for the chimera state in column (b), and $(0.4554, 0.04452, 0.0478)$ for the disordered state in column (c).}
 \label{fig:9fourier}
\end{figure*}

To numerically integrate the equations of motion we set initial conditions $\delta$-close to $(u^*, v^*) = (0,0)$ and we then use the {\em tsit5} solver (5th order adaptive step size
Runge Kutta method~\cite{Tsitouras2011}) to obtain time series in the time interval $[0,1000]$, every $0.01$ time units. More precisely, we set $\delta = 0.001$ and we draw uniform random numbers $\Delta u_i\in U[-1,1]$ and, similarly, for $\Delta v_j$, to eventually set 
\begin{align*}
    u_{i}(0) &= u^* + \delta \Delta u_i \quad i = 1,\ldots, \textcolor{black}{N_0}\\
    v_{j}(0) &= v^* + \delta \Delta v_j \quad j = 1,\ldots, \textcolor{black}{N_1}\, .
\end{align*}%
Based on some preliminary study, we observe that the system exhibits an oscillatory behavior, that can be regular or irregular, both in time and in space. In Fig.~\ref{fig:9dyn}, we report three typical cases for a ring of $\textcolor{black}{N_0}=50$ nodes and parameters $\textcolor{black}{\epsilon} = 0.05$, $b = 0.5$ and $c = -0.01$; in  panel (a) we can observe a regular state in which the system behaves as a traveling wave ($Q=P=9$), in panel (b) we display a chimera state where regular oscillations coexist with irregular behavior ($Q=P=12$), in panel (c) we show a disordered state where, i.e., no regular oscillations can be detected ($Q=P=21$). 
To quantify these dynamics, we propose a method rooted on the extraction of information about the instantaneous phases, amplitudes and frequencies from the time series obtained by numerically solving the dynamical equations. {\color{black}The method will be presented in the following section (we refer the interested reader to Appendix~\ref{sec:Fourier} for more details).
\subsection{The Fourier method}
\label{ssec:Fourier}
The Fourier transform can be used to extract information from signals. In particular, it enables the computation of quantities that are effectively ``local'' in time, such as amplitudes, frequencies, and phases. By restricting attention to the dominant amplitude within a given time window, any sufficiently regular signal $y(t)$ can be approximated as
\begin{align*}
   y(t)\sim a^{(w)}_0 + a^{(w)} \mathrm e^{\mathrm i(2\pi \Omega^{(w)} t + \theta^{(w)})}\,, 
\end{align*}
where $a^{(w)}_0 \in \mathbb{R}$ represents the baseline level, $a^{(w)} \in \mathbb{R}_+$ is the positive amplitude, $\Omega^{(w)} \in \mathbb{R}_+$  is the frequency, and $\theta^{(w)} \in [-\pi,\pi)$ is the phase. The superscript ($w$) indicates that these quantities are defined within the time window $w$, over which the approximation is assumed to hold true. Notably, if the signal is strictly periodic, these quantities are independent of the chosen window; therefore, by examining their variation across adjacent windows, one can infer the degree of regularity of the signal.

It is well known that the accuracy of the reconstructed amplitude, frequency, and phase obtained from the Fourier transform strongly depends on the length of the signal under consideration; in general, longer time windows reduce the reconstruction errors \textcolor{black}{at the cost of a ``poor'' time localization, because of the time-frequency uncertainty principle}. However, in the present framework, the signals are not necessarily stationary, and relatively short time windows are required to capture local variations in amplitude, frequency, and phase. To balance these competing requirements, we devise a modified Fourier method consisting of three steps.

First of all, we compute the Fast Fourier Transform of the signal $y(t)-\langle y\rangle$ on a given time window, $w=[t_0,t_1]$, where $\langle y\rangle$ is the time average of $y(t)$ in the time window. Let us assume the latter to be large enough to contain $n_{peaks}$ maxima, \textcolor{black}{meaning that the size of the time window is not fixed a priori, but based on the signal behavior. This allows us to have a relatively good frequency determination and, at the same time, a more precise temporal localization}. In this way, we obtain a first approximation of amplitude, frequency, and phase, respectively $\tilde{a}^{(w)}$, $\tilde{\Omega}^{(w)}$, and $\tilde{\theta}^{(w)}$. Hence, we also get the baseline $\tilde{a}^{(w)}_0=\langle y\rangle$. 

These values may, in principle, lack precision due to the finite size of the window. To obtain a more accurate estimate of ${\Omega}^{(w)}$, we exploit the fact that the amplitude of the FFT power spectrum is approximately quadratic in the vicinity of its maximum. By performing a quadratic fit in this region, we obtain improved estimates of the amplitude and frequency, denoted by $\hat{a}^{(w)}$ and $\hat{\Omega}^{(w)}$, respectively. \textcolor{black}{Let us observe that, in this way, we are able to circumvent the time-frequency uncertainty principle by correcting the frequency estimate obtained with the FFT.}

The third and last approximation is based on a nonlinear fit of the signal $y(t)$ of the form 
\begin{align*}
    \tilde{y}(t)=p_1\cos\left(2\pi\hat{\Omega}^{(w)} t+p_2\right)+p_3\, ,
\end{align*}
where we want to determine the unknown amplitude, $p_1$, phase, $p_2$, and baseline oscillation, $p_3$, by assuming $\hat{\Omega}^{(w)}$ to be precise enough. The nonlinear fit is initialized with the values previously obtained for the desired quantities, i.e., $\hat{a}^{(w)}$, $\tilde{\theta}^{(w)}$, and $\tilde{a}^{(w)}_0$. With the last step, we have eventually computed $a^{(w)}_0$, $a^{(w)}$, and $\theta^{(w)}$, while $\Omega^{(w)}=\hat{\Omega}^{(w)}$.

We then consider another time window, $w'=[t'_0,t'_1]$ and we repeat the same construction to finally get $a^{(w')}_0$, $a^{(w')}$, $\theta^{(w')}$, and $\Omega^{(w')}$. To ensure slow variation of the computed values on adjacent time windows, we impose the two windows to overlap, i.e., $[t_0,t_1]\cap[t'_0,t'_1]\not = \emptyset$. More precisely, we ensure they share a given number of maxima, $n_{share}$.

In this way we construct a sequence of time windows, $w_i$, $i=1,\dots,q$, and for each window we compute baseline, amplitude, phase, and frequency, $a^{(w_i)}_0$, $a^{(w_i)}$, $\theta^{(w_i)}$, and $\Omega^{(w_i)}$. We eventually compute the average:
\begin{equation}
\langle Z\rangle  = \frac{1}{q}\sum_{i=1}^{q}Z^{(w_i)} \, ,
\label{eq:averages}
\end{equation}
and, to measure the variation of each quantity across the different windows, we compute the respective standard deviations
\begin{equation}
    \label{eq:stds}
  (\sigma_{Z})^2 = \frac{1}{q-1}\sum_{i=1}^{q}(Z^{(w_i)}-\langle Z\rangle)^2\, ,  
\end{equation}
where $Z\in\{a_0,a,\Omega,\theta\}$. Fig.~\ref{fig:9fourier} displays a few examples of those metrics computed from three characteristic signals, namely, regular, chimera, and irregular, associated with the dynamical behaviors shown in Fig.~\ref{fig:9dyn}. The blue dots represent the averages~\eqref{eq:averages}, while the size of the shaded light blue regions is given by the standard deviations~\eqref{eq:stds}. In Fig.~\ref{fig:FFT} of Appendix~\ref{sec:Fourier}, we report some numerical results about the reconstruction of phase, amplitude, and frequency of a synthetic signal exhibiting a chimera-like behavior.}

{\color{black}
\subsection{Total normalized variations}
\label{ssec:TNVar}
}
Eventually, to measure the spatial dependence of those quantities, we define the total (normalized) variation~\cite{muolo2024phase} for phases, amplitudes and frequencies. The total normalized variation is a measure of the {\color{black}spatial} local regularity of those quantities computed by considering differences among nearby points: 
\begin{align}  
    V_\theta & = \frac{1}{\pi {\color{black}{N_0}}}{\sum_{i=1}^{{\color{black}{N_0}}} \min\{v_{\theta_i},\; 2\pi - v_{\theta_i}\}} \\
    V_a  &= \frac{1}{{\color{black}{N_0}}}{\sum_{i=1}^{{\color{black}{N_0}}}|\langle a\rangle_{i+1}- \langle a\rangle_i| } \\
    V_\Omega & = \frac{1}{{\color{black}{N_0}}}{\sum_{i=1}^{{\color{black}{N_0}}}|\langle \Omega \rangle_{i+1}- \langle \Omega \rangle_i| }\, ,
\end{align}
with $v_{\theta_i} = |\langle \theta\rangle_{i+1}- \langle \theta \rangle_i|$. Let us also observe that indexes should be considered modulo ${\color{black}{N_0}}$, i.e., ${\color{black}{N_0}}+1\equiv 1$, because of the ring structure.

The smaller the variation, the smoother the signal and vice versa. Hence, the latter qualifies to be a useful metric to classify chimera patterns. Let us consider the results presented in Fig.~\ref{fig:9fourier}. Panels in the column (a) show a very regular behavior for the variations: the frequency is constant, the amplitude slowly oscillated in space, finally the phases clearly shown a linear dependence on the node position. We are thus facing to a wave whose amplitude is not constant across the ring. This behavior can thus be classified as regular one. The panels in the column (b) display a different behavior: some nodes show a constant frequency while other one a ``parabola--like'' dependence on the node index, this is the classical behavior observed for chimera states~\cite{kuramoto_batt,Abrams2004,gopal2014}. The amplitudes also exhibit a nontrivial but smooth dependence on the node index, whereas the phases behave in a more irregular way. Finally, panels in the column (c) display an irregular behavior in the three quantities, even if small in amplitude and frequency.

\subsection{\color{black}A clustering method to identify chimera states}
\label{ssec:classific}
By applying the proposed method to compute the variations of data shown in Fig.~\ref{fig:9fourier} we obtain the following values for $(V_\theta, V_a, V_\Omega)$: $(0.0683, 0.082, 0.0022)$ in the
case of ordered behavior (column (a)),  $(0.2301, 0.1110, 0.1087)$ for the chimera state (column (b)) and $(0.4554, 0.04452, 0.0478)$ for the disordered case (column (c)). Even if a clear trend is shown, with higher variations associated to more irregular behavior as already stated, it also appears that there is not a clear ``universal'' threshold allowing to separate the different cases. For this reason, we resort to a classification algorithm to extract hidden patterns in the data. More precisely, we compute $(V_\theta, V_a, V_\Omega)$ for all $Q=P$ in $\{2,\dots,24\}$ and for each $P$ we repeat $20$ times the simulation. We thus obtain a database composed by $460$ triplets $(V_\theta, V_a, V_\Omega)$ {\color{black}(see panel (a) of Fig.~\ref{fig:hvsnmin:orient:13D} for a 3D visualization). \textcolor{black}{A preliminary analysis of the dataset allows us to conclude that the points are not distributed in well-separated spheres and, on the other hand, they exhibit nested clusters. For this reason, we resort to a hierarchical clustering. We then} employ a bottom-up approach to form the clusters by using the agglomerative clustering algorithm~\cite{TOKUDA2022126433}, building a tree known as a dendrogram  ("dendron" - "tree" and "gramma" -"draw")~\cite{Nielsen2016Hierachical}. We first assign all the variables to the leaves; then the method merges them into subsets to form larger clusters based on a notion of (dis)similarity (i.e., distance). Finally, we end up with a single cluster---the root. We define the \emph{depth threshold} $d$, also called similarity or height in the literature, which measures the distance between two nearby subsets. \textcolor{black}{Let us observe that, if the data were distributed in well-separated spheres, one could have used a different clustering technique, namely, one that can be partitional such as the K-means, where groups are formed by minimizing the weighted sum of the intra-cluster variances~\cite{Nielsen2016Kmeans}.}


We preprocess by rescaling in $[0,1]$ by using the Python function {\em MinMaxscaler}. The use of the agglomerative clustering algorithm allows us to construct a dendrogram of the given data (see panel (a) of Fig.~\ref{fig:hvsnmin:orient:1}). For the chosen value of the {\em depth threshold}, $d$, we observe three, well-separated, classes that, interestingly enough, correspond to the three typical behaviors above emphasized: regular, chimera, and irregular states. In the panel (b) of Fig.~\ref{fig:hvsnmin:orient:1}, we present a 2D projection of the 3D database, where we used only the coordinates $(V_\theta,V_a)$; let us observe that we here used the original values for the variations, i.e., prior to the pre-processing. Three clusters are clearly visible; the red cluster (class 1) corresponds to the ordered states associated with smaller variations, the blue cluster (class 2) corresponds to chimeras, showing larger variations in amplitudes and medium variations in phases, and the black cluster (class 3) corresponds to the disordered states, with larger variations in phases and medium variations in amplitude. In Fig.~\ref{fig:hvsnmin:orient:13D} we will present a complete 3D view of the data (panel (a)) together with the projections on the planes $(V_\theta,V_\Omega)$ (panel (b)) and $(V_a,V_\Omega)$ (panel (c)). 

With this projection, but also in the construction of the dendogram, we lose information about the used value of $Q=P$. Hence, for each value of $Q=P\in\{2,\dots,24\}$, we computed the statistical mode to identify the most frequent class among the $20$ realizations. Let us observe that this step is necessary because, for some values of $Q=P$, different dynamical behaviors can be observed. The results are reported in the panel (c) of Fig.~\ref{fig:hvsnmin:orient:1} and we can obverse the presence of ordered states for $Q=P$ ranging from $2$ to $10$, chimera states for the $Q=P\in\{11,12,13\}$, and then disordered states for larger values of $Q=P$, but for the values $\{16,17,18\}$ associated to regular behavior. {\color{black}
In Appendix~\ref{sec:compS}, we present results obtained using the SI measure~\cite{gopal2014}. We show that this measure is also capable of identifying the three classes, as is the proposed method. However, we further demonstrate that the SI-based classification is highly sensitive to parameter tuning.
}

Let us note that lowering the depth threshold causes the dendrogram to first identify four classes and then five (see Fig.~\ref{fig:orient1_5classes}). In the former case, the irregular class splits into two new classes, whereas in the latter case, the regular class also splits into two. In all cases, however, the chimera class maintains its integrity. {\color{black}It is worth noting that, if the system had exhibited different types of chimera states, this method could have been applied using additional data to identify finer classifications within the chimera class.}
\begin{figure*}[!ht]
\centering
\includegraphics[scale=0.145]{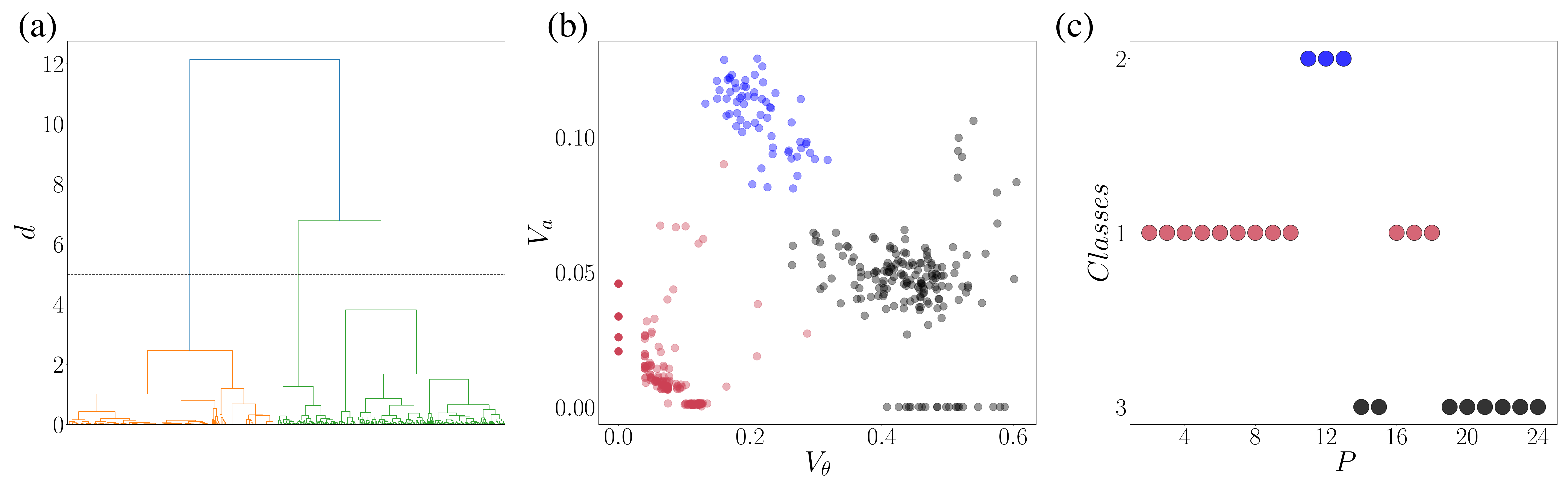}
\caption{\textbf{Classification of dynamical behaviors by using the variations $V_\theta$, $V_a$ and $V_\Omega$, in the case of orientation $1$ and $Q=P$ links have been reoriented.} In panel (a) we report the dendrogram obtained from the hierarchical clustering by using the values of  $(V_\theta, V_a, V_\Omega)$. Branch lengths represent inter-cluster distances prior to merging, indicating two well-separated groups. The horizontal black line denotes the depth threshold used to determine the number of classes. In panel (b) we report the projection in the plane $(V_\theta,V_a)$ of the three obtained clusters by using the agglomerative clustering: the red cluster (class 1) corresponds to the ordered states, the blue cluster (class 2) correspond to chimeras, and the black cluster (class 3) corresponds to the disordered states. Panel (c) shows an alternative view of the classification as a function of $P=Q$ in the range $\{2,\dots, 24\}$; to each value of $P=Q$ we associate the dominant class, i.e., determined from the statistical modal cluster membership of $(V_\theta, V_a, V_\Omega)$ and we can obverse the presence of ordered states for $Q=P$ ranging from $2$ to $10$, chimera states for the $Q=P\in\{11,12,13\}$, and then disordered states for larger values of $Q=P$, but for the values $\{16,17,18\}$.}
\label{fig:hvsnmin:orient:1}
\end{figure*}

We then consider the more general case where $P$ and $Q$ vary separately. More precisely for each $P\in\{3,\dots,24\}$ we let $Q\in\{0,\dots, P\}$ and for each couple $(P,Q)$ we consider $20$ realizations by randomly changing initial conditions. By following the same procedure presented above, we gather $6380$ triplets $(V_\theta, V_a, V_\Omega)$ and we apply the same classification algorithm. Three main classes are again found, corresponding to ordered, chimera, and disordered states, see panel (a) of Fig.~\ref{fig:hvsnmin:orient:1tri}; in  panel (b), we show a projection in the plane $(V_\theta, V_a)$. Finally, in panel (c), we show, for each value of $(P,Q)$, the associated class, defined again by using the statistical mode of the $20$ realizations; black points correspond to the disordered state, blue points to chimera behavior, and red points to the regular one. In addition to the $P$ values for which we found chimera in the case $P=Q$, we here observe chimera also for $P=10$ and $Q= 4,5,6,7$. 
\begin{figure*}[!ht]
\centering
\includegraphics[scale=0.145]{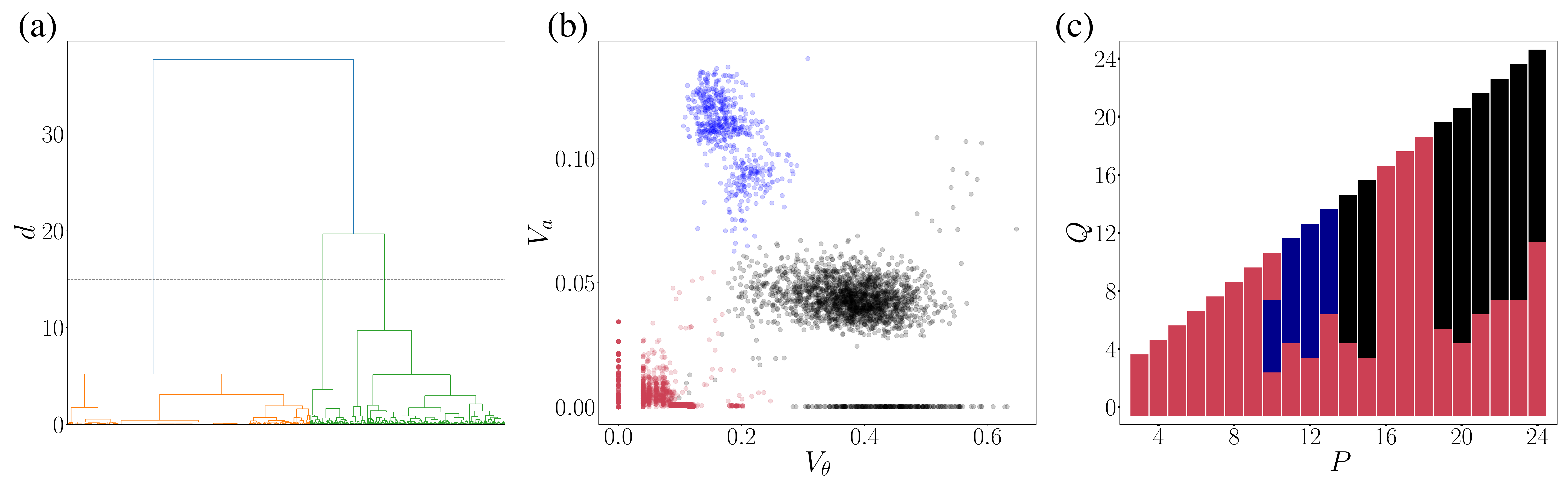}
\caption{\textbf{Classification of dynamical behaviors by using the variations $V_\theta$, $V_a$ and $V_\Omega$, in the case of orientation $1$ and $P$, $Q$ links have been reoriented with $0\leq Q\leq P$.} In panel (a) we show the dendrogram obtained from hierarchical clustering of $(V_\theta, V_a, V_\Omega)$; branch lengths represent inter-cluster distances prior to merging, indicating two well-separated groups. Panel (b) displays a 2D projection in the plane $V_\theta$, $V_a$ of the clusters so far obtained; points have been colored according to their class: 
 class 1 (red) associated to the ordered states, class 2 (blue) to denote chimeras and class 3 (black) corresponds to the disordered states. In panel (c) we report the classification as a function of $P$ and $Q$. For each value of the latter the dominant class is determined from the modal cluster membership of $(V_\theta, V_a, V_\Omega)$.}
\label{fig:hvsnmin:orient:1tri}
\end{figure*}

The results presented so far have been obtained for a fixed set of model parameters, $\textcolor{black}{\epsilon}=0.05$, $b=0.5$, and $c=-0.01$. Our aim is to briefly study the effects of those parameters on the dynamics. We thus fix again a ring made of $\textcolor{black}{N_0}=50$ nodes, and select three values of $P$ (we are here considering the case $Q=P$), each one corresponding to a main dynamical behavior, i.e., regular ($P=6$), chimera ($P=12$) and irregular ($P=21$), for the above fixed values of $\epsilon$, $b$ and $c$. We then vary $\epsilon$ and $b$, and we determine the system outcome; let us observe that to realize this last step, we used the classification method described above. More precisely, for values of $b\in[0.5,3.3]$, we numerically determine the time evolution of $u_i(t)$ and $v_j(t)$, we apply the Fourier method to compute $\langle a_i\rangle$, $\langle \Omega_i\rangle$ and $\langle \theta_i\rangle$, and then we compute the normalized variations. Eventually, we look at which class the triplet $(V_a,V_{\Omega},V_{\theta})$ belongs to, to conclude about the dynamical behavior. The results are reported in Fig.~\ref{fig:hvsnmin:orient:1b}, where we show, for ease of visualization, the projection in the $(V_{\theta},V_a)$ plane where the background shows with light colors the classes obtained with fixed $b=0.5$. Panel (a) corresponds to the choice $P=6$; for each $b$, we show the values of $(V_{\theta},V_a)$ corresponding to the statistical mode computed from $10$ replicas, as colored diamonds according to the value of $b$. One can observe that all the points lie in the region corresponding to the regular behavior. Stated differently, for all values of $b$, the system exhibits a regular behavior. Panel (b) displays the case $P=12$, by using the same ideas we report the values of $(V_{\theta},V_a)$ computed for each $b$ as colored diamonds, the darker the larger the value of $b$; all the \textcolor{black}{points belong to the class associated to chimera states, except one, which lies in the regular class.} Finally, panel (c) presents the case $P=21$; the values of $(V_{\theta},V_a)$ lie in the class associated with irregular behavior for all $b$. In conclusion, we have shown that by changing $b\in[0.5,3.3]$ the system almost never changes dynamical behavior showing a sort of robustness of system behavior with respect to changes in the parameter $b$. Let us observe that, besides the latter observation, having an ``atlas'' in the variations planes allows for a global understanding of the system behavior and, in particular, of possible changes of the latter.
\begin{figure*}[!ht]
\includegraphics[scale=0.14]{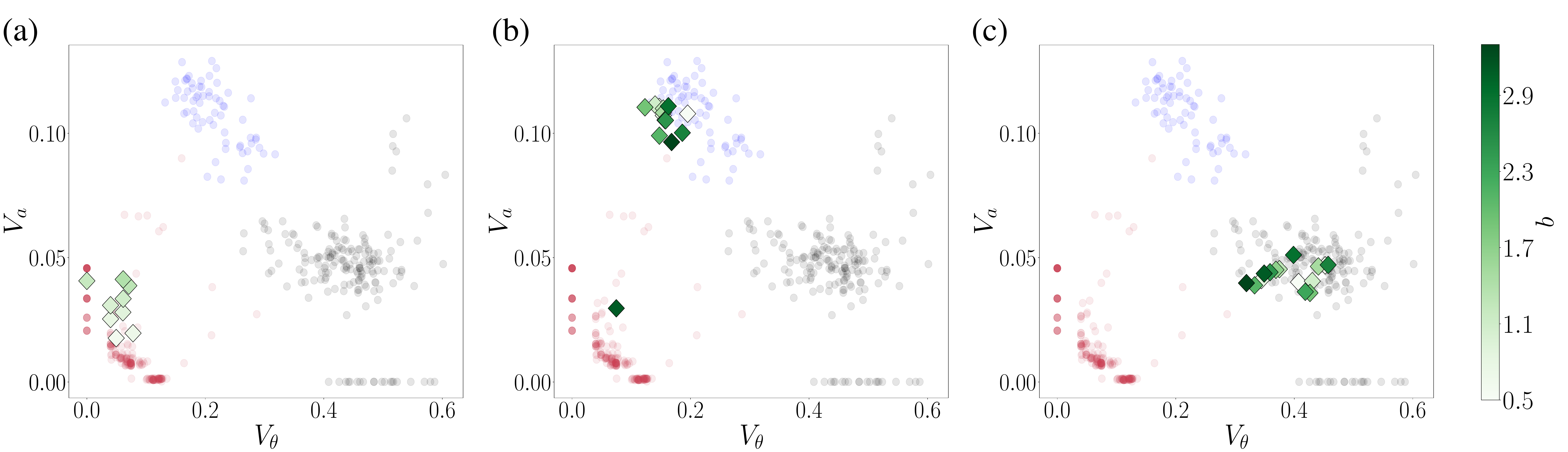}
\caption{\textbf{Impact of the parameter $b$ on system dynamics}. For values of the parameter $b$ in \textcolor{black}{$[0.5,3.3]$} we compute the normalized total variations associated to the Fourier metrics for the orbits. We then report the values $(V_\theta,V_a)$ as diamonds in the plane $(V_\theta,V_a)$ containing in background also the classes obtained so far with fixed $b=0.5$. Diamonds are colored according to the value of $b$, the larger the value the darker the color. Panel (a) corresponds to $Q=P=6$, we can observe that all the values of $(V_\theta,V_a)$ lie in the same class, that corresponds to regular behavior. Panel (b) shows the case $Q=P=12$, \textcolor{black}{we can observe that, by increasing $b$, the system can move from chimera to regular (see the dark diamond in the bottom left corner)}. Panel (c) displays the case $Q=P=21$, one more time the points belong to a single class, the irregular one.}
\label{fig:hvsnmin:orient:1b}
\end{figure*}

We then repeat a similar analysis by varying $\epsilon\in[0.01,0.1]$. The results are shown in Fig.~\ref{fig:hvsnmin:orient:1a} by applying the same ideas of the previous case. In panel (a) ($P=6$) and panel (c) ($P=21$), we can observe that the couples $(V_{\theta},V_a)$, again reported as diamond colored according to the value of $\epsilon$, remain in the same class, i.e., the system exhibits the same behavior. A \textcolor{black}{phenomenon similar to the one already discussed in Fig.~\ref{fig:hvsnmin:orient:1b}} is visible in panel (b)($P=12$): we can observe that points are spread among two classes, the regular one for small values of $\epsilon$, i.e., lighter color, and the chimera one for larger values of $\epsilon$, i.e., darker color. The transition value being about $\epsilon\sim 0.03$.
\begin{figure*}[!ht]
\includegraphics[scale=0.14]{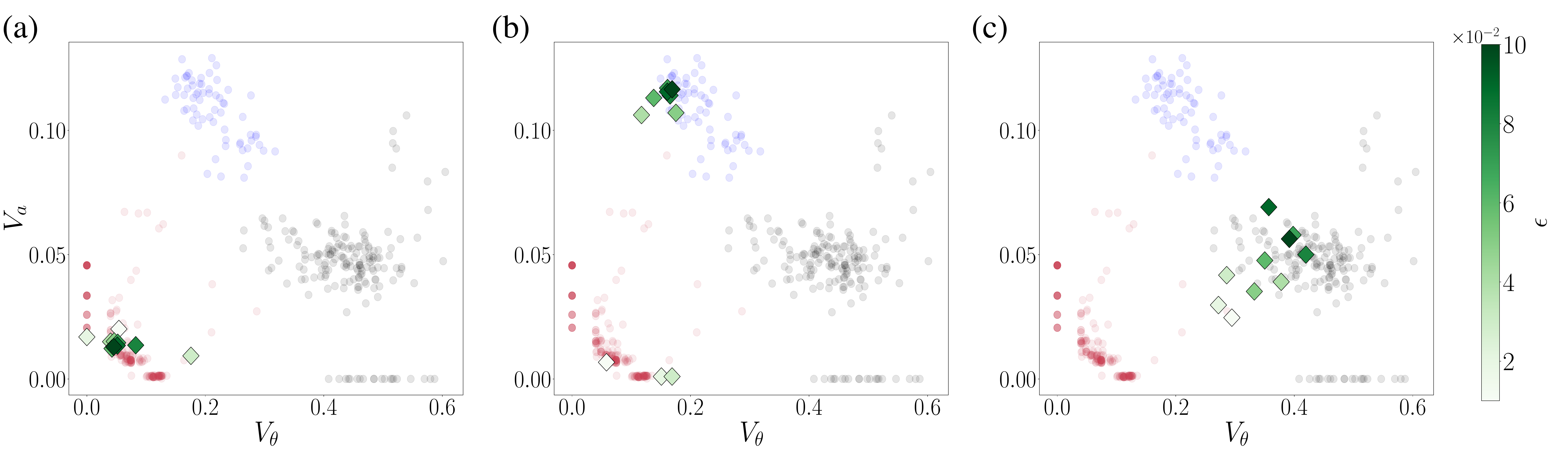}
\caption{\textbf{Impact of the parameter $\textcolor{black}{\epsilon}$ on system dynamics}. For values of the parameter $\textcolor{black}{\epsilon}$ in $[0.01,0.1]$ we compute the normalized total variations associated to the Fourier metrics for the orbits. We then report the values $(V_\theta,V_a)$ as diamonds in the plane $(V_\theta,V_a)$ containing in background also the classes obtained so far with fixed $\textcolor{black}{\epsilon}=0.05$. Diamonds are colored according to the value of $\textcolor{black}{\epsilon}$, the larger the value the darker the color. Panel (a) corresponds to $Q=P=6$, we can observe that all the values of $(V_\theta,V_a)$ lie in the same class, that corresponds to regular behavior. Panel (b) shows the case $Q=P=12$. Here, light diamonds, i.e., associated to small values of $\textcolor{black}{\epsilon}$, remain in the regular class, while darker diamonds, i.e., for larger values of $\textcolor{black}{\epsilon}$, fall in the chimera states. Panel (c) displays the case $Q=P=21$; one more time the points belong to a single class, the irregular one.}
\label{fig:hvsnmin:orient:1a}
\end{figure*}
\\
\section{Conclusion}
\label{sec:concl}

In this work, we introduced a method to \textcolor{black}{identify} chimera states based on signal features extracted from time series. Starting from the simulated dynamics, i.e., the time series of each node, we computed instantaneous phases, amplitudes, and frequencies via a modified Fourier method, relying on the use of windowed data. Then, we evaluated their total normalized variations to obtain a measure of local regularity of the computed quantities. These indicators yield a compact representation of the spatial organization of the system. By embedding the latter in a three-dimensional space and by applying hierarchical clustering algorithms and dendrograms, we obtained a classification of the observed dynamical regimes without relying on ad hoc thresholds.

As an example, we considered the {\color{black}topological} FitzHugh–Nagumo model with dynamics defined on both nodes and links within the framework of topological signals. This setting provided a system with sufficiently rich behavior by including coherent, chimera, and irregular regimes. We analyzed two orientations for the links of the $1$-simplicial complex used to couple the variables. We then investigated the effects of the coupling range and the number of reoriented links to generate a variety of dynamical patterns on which to test the method. We then tested the proposed approach on the resulting time series and showed that it was able to distinguish between ordered, chimera, and disordered states. In particular, the method captured intermediate and weakly structured regimes that are usually difficult to identify by using standard measures. The classification remained stable across different parameter values and network configurations, indicating that the procedure is robust. {\color{black} In Appendix \ref{sec:compS}, we further corroborated the robustness of our method by using, instead of the topological FHN, a network of FHN oscillators anchored to the nodes of a ring and compared our identification method with previously existing ones, in particular, with the strength of incoherence.}

{\color{black}
In conclusion, in this work, we have introduced and discussed an identification method capable of clearly discriminating between different dynamical regimes, namely, regular, irregular, and chimera states. This allows us to answer the challenging question about the threshold value needed in previous methods, e.g., the strength of incoherence, to identify chimera states. The proposed method can be summarized in the following pipeline:
\begin{enumerate}
    \item Create time series from numerical simulations or obtain them from experiments;
    \item Perform the (modified) Fourier analysis to extract relevant information about phases, amplitudes, and frequency;
    \item Compute the (normalized) total variations to compact the information obtained from the previous point;
    \item Apply a statistical classification scheme to extract the dynamical features from the data.
\end{enumerate}
It is important to note that the proposed scheme extends beyond the present application to chimera states, both in terms of the underlying model and the research question being addressed. Furthermore, the effectiveness of the method stems primarily from the sequence of processing steps rather than from any specific implementation of those steps. Each stage can be realized with considerable flexibility. For example, the system dynamics may be obtained using any suitable numerical integration scheme or from experimental measurements. Likewise, alternative techniques, such as wavelet-based methods, may be employed to extract information about the oscillatory behavior of the trajectories. For the classification stage, other approaches may also be adopted, depending on the characteristics of the data and the outcome of a preliminary analysis of its distribution.
}

Finally, since the method relies only on signal-based quantities and not on specific properties of the underlying model, it is not restricted to the framework considered here, but can be directly applied to other dynamical systems, as well as to experimental or real-world data, whenever a systematic identification of different dynamical regimes is required.

\acknowledgments

The work of S.N.J. is supported by the Institutional call for doctoral fellowship (CERUNA), University of Namur. 
R.M. acknowledges JSPS KAKENHI 24KF0211 for financial support.
The work of P.M. is supported by the Ministry of Education-Rashtriya Uchchatar Shiksha Abhiyan (MoE RUSA 2.0): Bharathidasan University -- Physical Sciences. 

\section*{Author's contribution} 
S.N.J.: software,  methodology, investigation, visualization, formal analysis, validation, writing -- original draft, writing -- review and editing. R.M.: conceptualization, supervision, writing -- original draft, writing -- review and editing. P.M. supervision. T.C.: conceptualization, methodology, investigation, visualization, formal analysis, supervision, writing -- original draft, writing -- review and editing. All authors read and approved the manuscript.


\bibliography{reference}

\onecolumngrid
\appendix

{\color{black}
\section{Stability of the homogeneous equilibrium}
\label{sec:stabhomeq}

The aim of this section is to prove the linear stability of the the equilibria of system given by Eqs.~\eqref{eq:eq1aprime} and~\eqref{eq:eq1bprime}, hereby rewritten sake of simplicity
\begin{align*}
\frac{d{u}_i}{dt} & = \textcolor{black}{\epsilon'}(u_i - \frac{u_i^3}{3}) - \textcolor{black}{\epsilon'}\sum_lB_{1}[i,l]v_l
 \\ 
\frac{dv_j}{dt} & = \sum_iB_{1}[i,j]u_i + b + cv_j 
\end{align*}
where $\textcolor{black}{\epsilon'} = 1/\textcolor{black}{\epsilon}$. To determine the stability of the homogeneous solution $(u_i,v_j)=(u^*,v^*)$, where $(u^*,v^*)=(\pm\sqrt{3},-b/c)$ or $(u^*,v^*)=(0,-b/c)$, we consider the perturbations, $\delta u_i = u_i - u^*$ and $\delta v_j = v_j - v^*$, we insert them into the above equations and then we linearize to get
\begin{align}
 \frac{d\delta u_i}{dt} & = \textcolor{black}{\epsilon'}(1 - (u^*)^2)\delta u_i - \textcolor{black}{\epsilon'} \sum_lB_{1}[i,l] \delta v_l\label{eq:2a}\\ 
 \frac{d\delta v_j}{dt} & = \sum_iB_{1}[i,j] \delta u_i + c\delta v_j \label{eq:2b}\, ,
\end{align}
or in matrix form
\begin{align*}
 \frac{d}{dt} \begin{pmatrix}
 \delta u \\
 \delta v
 \end{pmatrix} 
 =
 \mathbf{J} \begin{pmatrix}
 \delta u \\
 \delta v
 \end{pmatrix},
\end{align*}
where we introduced the $(N_0 + N_1) \times (N_0 + N_1) $ Jacobian matrix $\mathbf{J} = \begin{bmatrix} \textcolor{black}{\epsilon'}(1 - (u^*)^2)\mathbf{I}_{N_0} & -\textcolor{black}{\epsilon'}\mathbf{B}_1 \\
\mathbf{B}_1^\top & c\mathbf{I}_{N_1}
\end{bmatrix}$ and the stack vectors $\delta u=(\delta u_1,\dots,\delta u_{N_0})^\top$ and $\delta v=(\delta v_1,\dots,\delta v_{N_1})^\top$. To gain some analytical insight into the problem, let us reduce the dimension of the latter by using $\mathbf{B}_1 \mathbf{B}_1^\top = \mathbf{L}_0$, i.e., the network Laplace matrix, and $\mathbf{B}_1^\top\mathbf{B}_1 = \mathbf{L}_1$, the $1$-Hodge-Laplace matrix. The eigenvalues of $\mathbf{L}_0$ and $\mathbf{L}_1$ are $\Lambda^{(\alpha)}_0= \Lambda^{(\alpha)}_1$ and they can be expressed as the square of the singular value $b_\alpha$ of the matrix $\mathbf{B}_1$, namely $\Lambda^{(\alpha)}_0 =\Lambda^{(\alpha)}_1 = b_\alpha^2$. On the other hand the eigenvectors $\psi^\alpha_0$ and $\psi^\alpha_1$ obey 
\begin{align}
\label{eq:phi1}
 \mathbf{B}_1\psi_\alpha^1 & = b_\alpha\psi_\alpha^0 \\
 \label{eq:phi0}
 \mathbf{B}_1^\top\psi_\alpha^0 & = b_\alpha\psi_\alpha^1 \, .
\end{align}
We can now project $\delta u_i$ and $\delta v_j$ onto the eigenbasis: $\delta u_i = \sum_{\alpha} \delta \hat{u}_\alpha (\psi_{\alpha}^0)_i$ and $\delta v_j = \sum_\alpha \delta \hat{v}_\alpha (\psi_\alpha^1)_j$. By using Eqs.~\eqref{eq:phi1} and~\eqref{eq:phi0} we then get from Eqs.~\eqref{eq:2a} and~\eqref{eq:2b}
\begin{align*}
 \frac{d\delta {\hat{u}}_\alpha}{dt} & = \textcolor{black}{\epsilon'}(1 - (u^*)^2) \delta \hat{u}_\alpha - \textcolor{black}{\epsilon'} b_\alpha\delta \hat{v}_\alpha\\
  \frac{d\delta {\hat{v}}_\alpha}{dt} & = c  \delta \hat{v}_\alpha +  b_\alpha \delta \hat{u}_\alpha
\end{align*}
or in matrix form
\begin{align*}
 \frac{d}{dt} \begin{pmatrix}
 \delta {\hat{u}}_\alpha\\
 \delta {\hat{v}}_\alpha
 \end{pmatrix} 
 =
 \begin{pmatrix}
 \textcolor{black}{\epsilon'}(1 - (u^*)^2) & - \textcolor{black}{\epsilon'} b_\alpha \\
 b_\alpha & c
 \end{pmatrix} 
 \begin{pmatrix}
 \delta {\hat{u}}_\alpha\\
 \delta {\hat{v}}_\alpha
 \end{pmatrix} =:\mathbf{J}_\alpha  \begin{pmatrix}
 \delta {\hat{u}}_\alpha\\
 \delta {\hat{v}}_\alpha
 \end{pmatrix} 
\end{align*}
The stability of the homogeneous solution $(u^*,v^*)$ can be determined by studying the spectrum of $\mathbf{J}_\alpha$. A straightforward computation returns the roots of the characteristic polynomial
\begin{align*}
\lambda = \frac{\textcolor{black}{\epsilon'}u' + c \pm \sqrt{(\textcolor{black}{\epsilon'}u' + c)^2 - 4(c\textcolor{black}{\epsilon'}u'+\textcolor{black}{\epsilon'}b_\alpha^2)}}{2}, 
\end{align*}
where $u' = (1- (u^*)^2)$. By direct inspection of the latter relation, one can easily infer about the stability of the equilibria.
}

\section{The Fourier method}
\label{sec:Fourier}
{In the main text, Section~\ref{ssec:Fourier}, we presented the method based on the Fourier transform, used to extract amplitude, phase and frequency, from the signals obtained from a numerical integration of Eqs.~\eqref{eq:eq1a} and~\eqref{eq:eq1b}. The aim of this section is to provide some additional details and show the effectiveness of the method on a synthetic signal, roughly inspired form the chimera state reported in the middle panel of Fig.~\ref{fig:9dyn}. \textcolor{black}{Again, let us stress that the use of an improved Fourier method is required by the ``shape'' of the signals, whose frequency can depend on the spatial index. We must, thus, rely on a method that, at the same time, allows to have a good estimate for frequency, amplitude, and phase, but also to be time localized, since these quantities can vary in time.}

Given a sufficiently regular signal $y(t)$ on a time window $w=[t_0,t_1]$, we can approximate it by
\begin{align*}
   y(t)\sim a^{(w)}_0 + a^{(w)} \mathrm e^{\mathrm i(2\pi \Omega^{(w)} t + \theta^{(w)})}\,, 
\end{align*}
being $a^{(w)}_0 \in \mathbb{R}$ the baseline level of oscillation, $a^{(w)} \in \mathbb{R}_+$ the positive amplitude of the oscillation, $\Omega^{(w)} \in \mathbb{R}_+$  the frequency, and $\theta^{(w)} \in [-\pi,\pi)$ the phase. For periodic signals, the latter quantities are constant and thus independent of time window, deviations from this behavior will thus provide information about the regularity of the signal.

By applying the discrete Fast Fourier Transform~\cite{FFT} to the signal $y(t_i)-\langle y\rangle$ evaluated for $t_i\in w=[t_0,t_1]$, where $\langle y\rangle$ is the time average of $y(t)$ in the time window, we obtain estimated values for amplitude, frequency and phase, respectively $\tilde{a}^{(w)}$, $\tilde{\Omega}^{(w)}$, and $\tilde{\theta}^{(w)}$. The latter are obtained by using the classical discrete FFT algorithm returning for $\tilde{\Omega}^{(w)}$ the frequency that maximizes the power spectrum of the signal and for the amplitude $\tilde{a}^{(w)}$ the module of the power spectrum. The spectrum is however known only at discrete values, as for the signal, and thus the time resolution and the window size impact the accuracy of $\tilde{a}^{(w)}$ and $\tilde{\Omega}^{(w)}$. By leveraging on the fact that the power spectrum has a parabola-like behavior close to its maximum, we can improve the previous estimates by interpolate the power spectrum with a quadratic polynomial close to the maximum, the amplitude will thus be given by the module of the maximum and the frequency by its position (see Fig.~\ref{fig:FFTsign}).
\begin{figure}[!ht]
\includegraphics[width=1\linewidth]{FigFFT.pdf}
\caption{{\color{black}\textbf{Use of the FFT and the quadratic interpolation to compute amplitude and frequency.} In panel (a) we display the signal $y(t)$ restricted to the time window $[950,952]$ where $n_{peak}=10$ are visible. In panel (b) we show a portion of the power spectrum \textcolor{black}{(blue dashed curve)} and the quadratic interpolation \textcolor{black}{(red curve)} close to the maximum of the power spectrum. Panel (c) present a zoom in the neighborhood of the maximum: the blue dot with coordinates $(3.9648,1.2485)$ is the one obtained by direct use of FFT, the black square with coordinates $(3.9722,1.2488)$ has been obtained as the maximum of the parabola.}}
\label{fig:FFTsign}
\end{figure}

Once the latter approximation for amplitude, phase, and frequency has been obtained, we look for a last improvement  based on a nonlinear fit of the signal $y(t)$ in the form 
\begin{align*}
    \tilde{y}(t)=p_1\cos\left(2\pi\hat{\Omega}^{(w)} t+p_2\right)+p_3\, ,
\end{align*}
the unknown being amplitude, $p_1$, phase, $p_2$, and baseline oscillation, $p_3$, by assuming $\hat{\Omega}^{(w)}$ to be precise enough, thanks to the previous estimate. The latter are also used as starting point for the nonlinear fit implemented via a least square nonlinear optimization algorithm~\cite{Matlab}.

We then consider $q$ time windows, $w_i$, large enough is such a way the signal $y(t)$ exhibits $n_{share}$ oscillations in the intersection between two consecutive windows. We obtain for $i=1,\dots,q$, the baseline, amplitude, phase, and frequency, $a^{(w_i)}_0$, $a^{(w_i)}$, $\theta^{(w_i)}$, and $\Omega^{(w_i)}$, and we eventually compute their averages~\eqref{eq:averages} and standard deviations~\eqref{eq:stds}.} Fig.~\ref{fig:9fourier} in the main text, displays few examples of those metrics computed from three characteristic signals, regular, chimera and irregular. The blue dots represent the averaged~\eqref{eq:averages} while the size of the shaded light blue regions are given by the standard deviations~\eqref{eq:stds}.

In Fig.~\ref{fig:FFT} we report some numerical results about the reconstruction of phase, amplitude and frequency of a given signal of the form
\begin{equation}
\label{eq:syntsig}
    x_j(t) = a_{0}(j)\cos\left[2\pi\Omega_{0}(j)t+\theta_{0}(j)\right]\quad \forall j=1,\dots,100\, .
\end{equation}
Where the phases depend linearly on the node index $j$, i.e., $\theta_0(j)=\pi (j-1)/100$ (see black squares in panel (a) of Fig.~\ref{fig:FFT}). The amplitude $a_0$ depends on the node index $j$ in a ``multi-bumps parabola--like'' way (see black squares in panel (b) of Fig.~\ref{fig:FFT}), i.e.,
\begin{equation*}
a_0(s)=1+20
\begin{cases}
        s(1/5-s) & \text{if $0 \leq s \leq 1/5$}\\
        (s-1/5)(2/5-s)& \text{if $1/5 < s \leq 2/5$}\\
        (s-2/5)(3/5-s)& \text{if $2/5 < s \leq 3/5$}\\
        (s-3/5)(4/5-s)& \text{if $3/5 < s \leq 4/5$}\\
        (s-4/5)(1-s)& \text{if $4/5 < s \leq 1$}
    \end{cases}
\end{equation*}
with $s=(j-1)/100$. Similarly the frequency exhibits two-bumps (see black squares in panel (c) of Fig.~\ref{fig:FFT})
\begin{equation*}
\Omega_0(s)=1+0.1
\begin{cases}
        (s-1/5)(2/5-s)& \text{if $1/5 \leq s \leq 2/5$}\\
        (s-3/5)(4/5-s)& \text{if $3/5 \leq s \leq 4/5$}\\
        0& \text{otherwise}
    \end{cases}
\end{equation*}
with $s=(j-1)/100$. Let us observe that such functional dependence has been chosen to mimic the behavior of $a_i$ and $\Omega_i$ shown in Fig.~\ref{fig:9fourier}, as we can appreciate by looking at the time space-time plot of the signal $x_j(t)$ (see Fig.~\ref{fig:reconstrsign}).
\begin{figure}[!ht]
\includegraphics[width=0.5\linewidth]{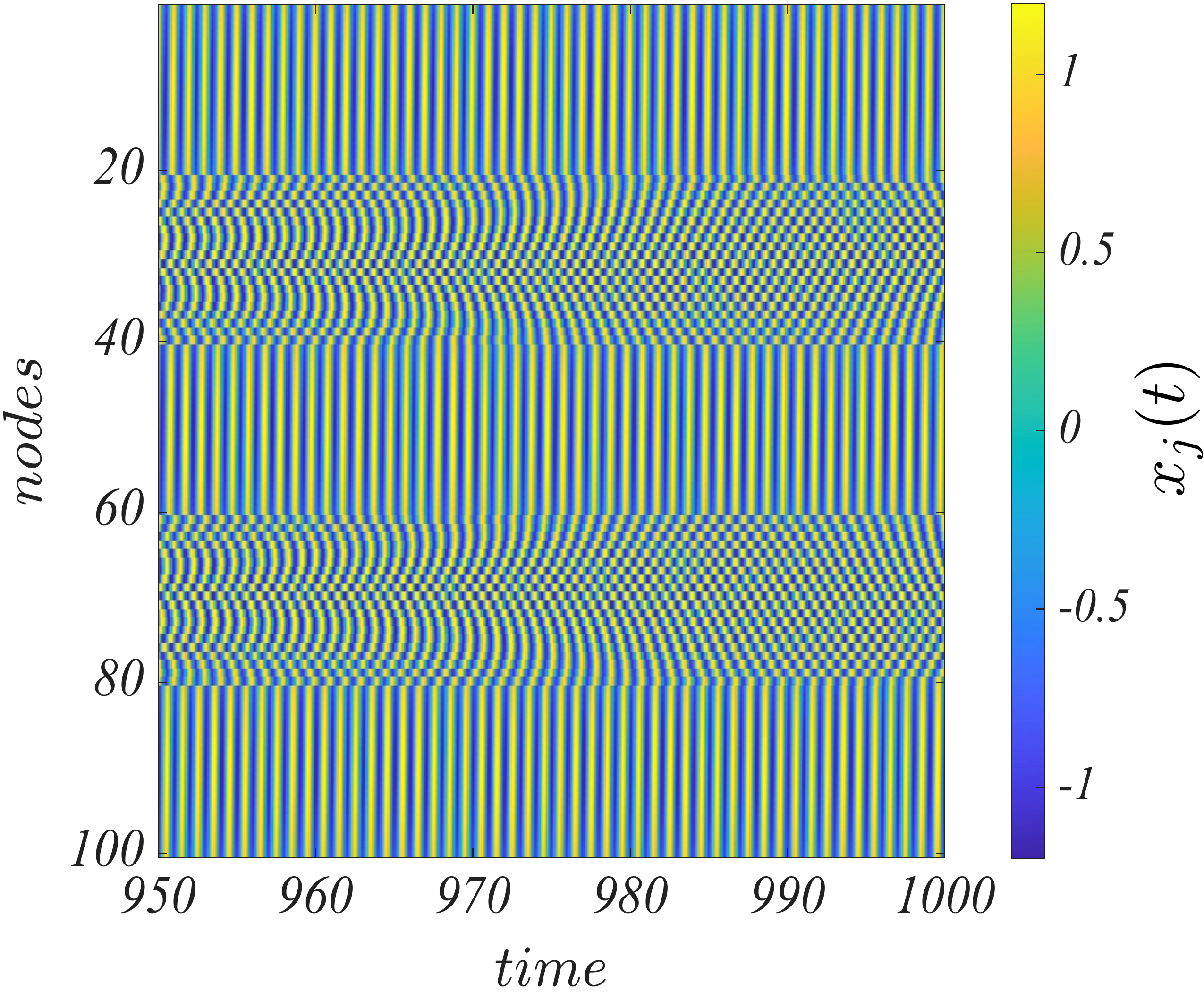}
\caption{\textbf{Synthetic signal used to show the application of the Fourier method}. We display the synthetic signal $x_j(t)$, built according to Eq.~\eqref{eq:syntsig} and whose reconstructed features are presented in Fig.~\ref{fig:FFT}.}
\label{fig:reconstrsign}
\end{figure}

From the knowledge of the signal $x_j(t)$ on the time interval $[500,1000]$ for all $j=1,\dots,100$, we reconstruct $\langle\theta_j\rangle$, $\langle a_j\rangle$ and $\langle\Omega_j\rangle$, by using $n_{peaks}=10$, i.e., by considering a time window containing $10$ oscillations, and $n_{share}=5$, i.e., by letting the consecutive time windows to overlap by $5$ oscillations. The results are reported in Fig.~\ref{fig:FFT} panels (a), (b) and (c), by using green dots superposed to the original values (black squares); we can observe that the agreement is very good as it can be confirmed by looking at the errors, i.e., $\langle \theta\rangle-\theta_0$ (panel (d)), $\langle a\rangle-a_0$ (panel (e)), and $\langle \Omega\rangle-\Omega_0$ (panel (f)).
\begin{figure*}[!ht]
\includegraphics[scale=0.145]{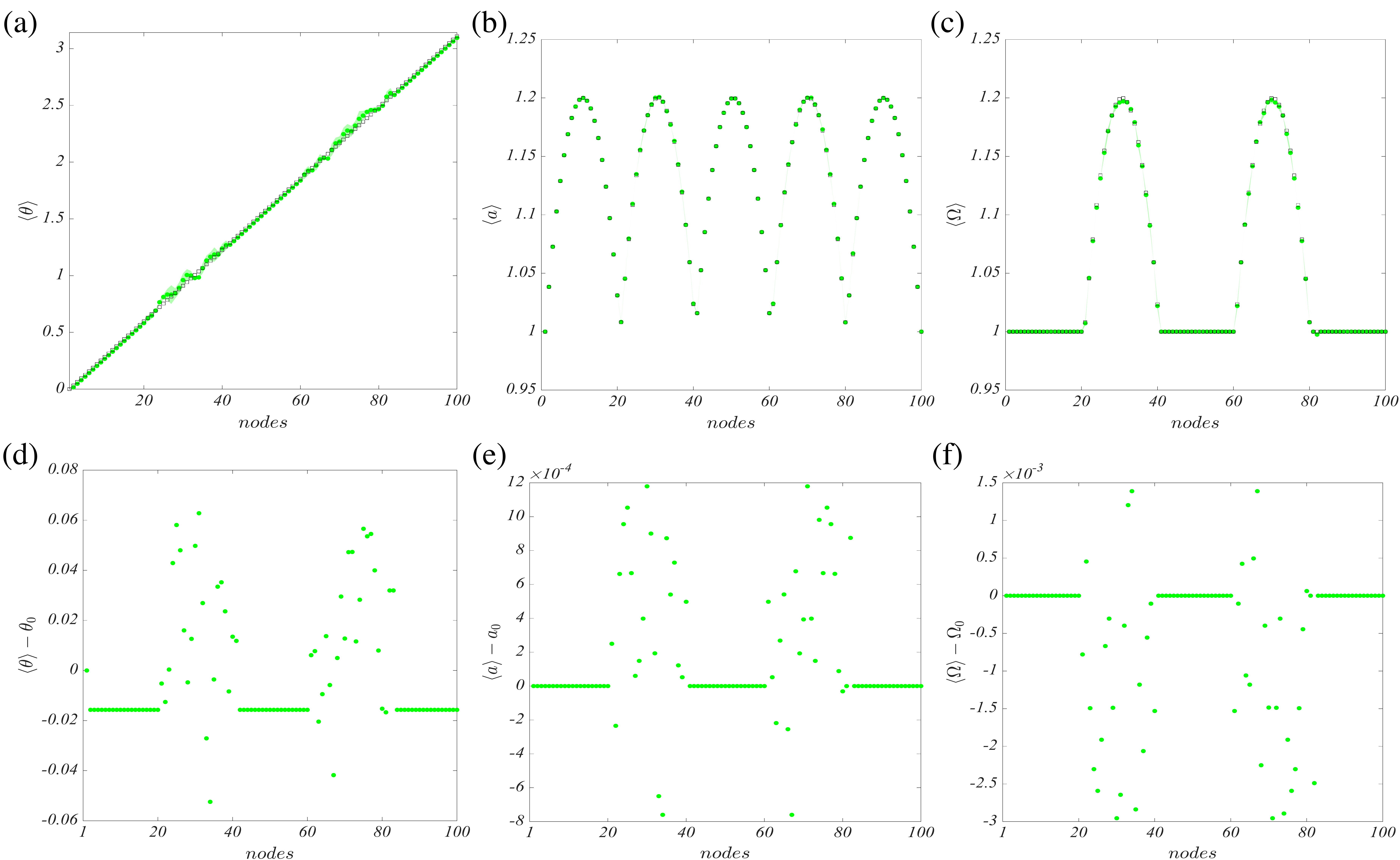}
\caption{\textbf{The Fourier method}. We present an application of the Fourier method to reconstruct phase, amplitude and frequency of a given signal. In panel (a) we report the original (black square) and reconstructed (green dots) phases, in panel (b) the original (black square) and reconstructed (green dots) amplitudes and in panel (c) the original (black square) and reconstructed (green dots) frequency. The bottom panels show the reconstruction error $\theta_0-\langle \theta\rangle$ (panel (d)), $a_0-\langle a\rangle$ (panel (e)), and $\Omega_0-\langle \Omega\rangle$ (panel (f)).}
\label{fig:FFT}
\end{figure*}

\section{More details about the dynamics}
\label{sec:moredetdyn}

Let us observe that in the case of orientation $1$ with $Q=P=1$, the dynamics does not show an oscillatory behavior, it moves away from the small perturbation of the initial homogeneous state, it (almost) reaches another homogeneous state but without any clear pattern in the studied time window (see Fig.~\ref{fig:dynamics_P=1} for a typical dynamical behavior), so our method cannot be applied to compute $\langle \theta\rangle$, $\langle a\rangle$ and $\langle \Omega\rangle$. A similar conclusion holds true for $P=2$ and $Q=0,1$. Hence, we present in Fig.~\ref{fig:hvsnmin:orient:1} the result of the analysis for $P\geq 2$, and $P\geq 3$ in Fig.~\ref{fig:hvsnmin:orient:1tri}.
\begin{figure}[!ht]
   \includegraphics[width=0.5\linewidth]{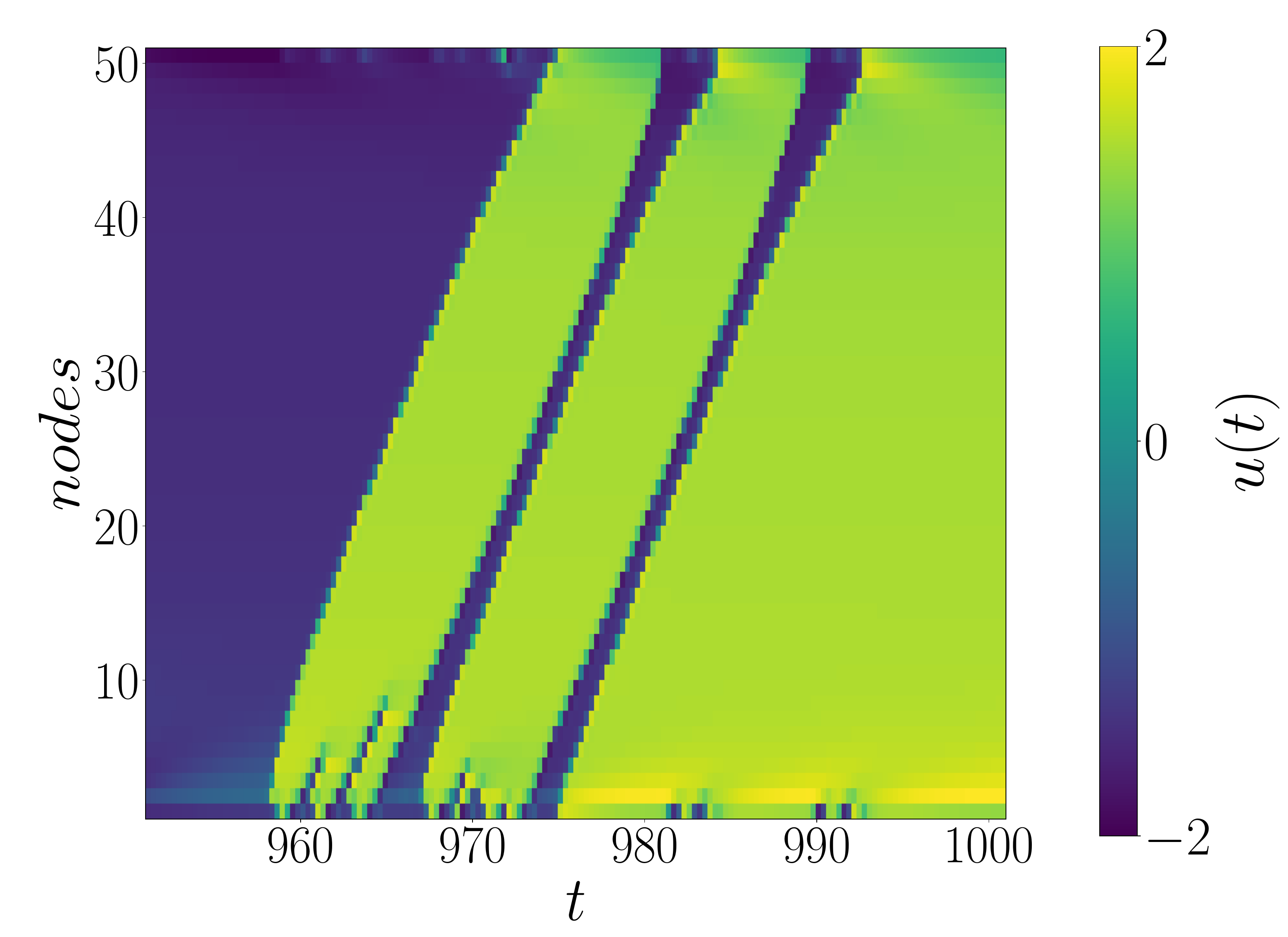}
   \caption{\textbf{Dynamical behavior in the case $P=1$ for orientation $1$}. We show the dynamical behavior of a typical numerical solution of the FHN system in the case of orientation $1$ with $P=Q=1$ reoriented links. One can observe that the system diverges from the initial almost homogeneous solution, but it does not seem to exhibit any oscillatory behavior in the considered time window.}
   \label{fig:dynamics_P=1}
\end{figure}

\section{More details about the classification}
\label{sec:moredetclass}

\textcolor{black}{Some other ways to represent the classification of dynamical behaviors by using the variations $V_\theta$, $V_a$, and $V_\Omega$, in the case of orientation $1$ and $Q=P$ links have been reoriented, are shown in Fig.~\ref{fig:hvsnmin:orient:13D}. In panel (a), we report a 3D view of the clusters obtained by using the agglomerative clustering applied to the values of  $(V_\theta, V_a, V_\Omega)$, with the depth threshold fixed at the same value used in Fig.~\ref{fig:hvsnmin:orient:1}. In panel (b), we report the projection in the plane $(V_\theta,V_\Omega)$, while, in panel (c), the 2D projection $(V_a,V_\Omega)$.}
\begin{figure*}[!ht]
\centering
\includegraphics[scale=0.14]{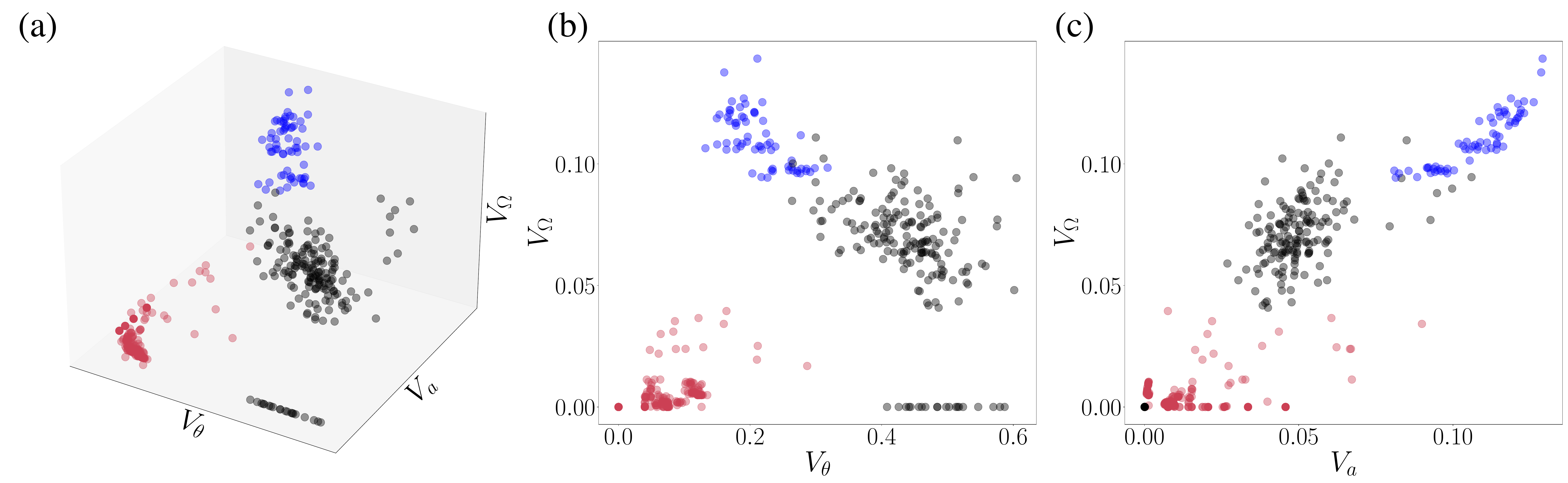}
\caption{\textbf{A complete view of the classification of dynamical behaviors by using the variations $V_\theta$, $V_a$ and $V_\Omega$, in the case of orientation $1$ and $Q=P$ links have been reoriented.} In panel (a) we report a 3D view of the clusters obtained by using the agglomerative clustering obtained by using the values of  $(V_\theta, V_a, V_\Omega)$, with the depth threshold fixed at the same value used in Fig.~\ref{fig:hvsnmin:orient:1}. In panel (b) we report the projection in the plane $(V_\theta,V_\Omega)$ while in panel (c) the 2D projection $(V_a,V_\Omega)$.}
\label{fig:hvsnmin:orient:13D}
\end{figure*}

\textcolor{black}{In Fig.~\ref{fig:hvsnmin:orient:1}, we have set a depth threshold returning three clusters, however we can obtain a finer classification by lowering the depth. From the dendrogram displayed in Fig.~\ref{fig:hvsnmin:orient:1}, one can observe that the next possible clustering is four where class 3 splits into two disordered classes. If we decrease the depth threshold even more, we obtain five clusters, now the ordered state also split into two clusters corresponds to two different regular behaviors, almost homogeneous and traveling waves. The fact that the chimera class does not split is due to the fact that the studied model exhibits a single kind of chimera behavior; on the other hand our classification method could be used to discriminate among different kinds of chimera states if they were present in the model and if enough data are provided.}
The results are shown in panel (a) of Fig.~\ref{fig:orient1_5classes}. In panel (b), we report the projection in the plane $(V_\theta,V_a)$ of the {\color{black} five} obtained clusters: the light red cluster (class 1,1) and dark red cluster (class 1,2) correspond to the ordered states, emerging from the unique regular class shown in Fig.~\ref{fig:hvsnmin:orient:1}, the blue cluster (class 2) corresponds to chimeras, and the black (class 3,1) and light black cluster (class 3,2) correspond to the disordered states. The new dynamical behaviors for the split classes are shown in Fig.~\ref{fig:orient1_otherdyn}. The top row correspond to the regular behavior while the bottom row correspond to the irregular one. The former can be classified as weakly homogeneous solution and (almost) twisted state, and they are shown for the cases $P=Q=3$ and $P=Q=7$ respectively. In the bottom row two disordered states are shown for the cases $P=Q=21$ and $P=Q=24$. The values of $(V_\theta, V_a, V_\Omega)$ are $(0.0000, 0.0336, 0.0000)$ for (a) showing weakly homogeneous solution , $(0.0747, 0.0070, 0.0019)$ for (b) exhibiting twisted states, $(0.4297, 0.0495, 0.0637)$ for (c)  with irregular behavior and $(0.4082,0.00005,0.00002)$ for (d) showing irregular behavior with very small variation in $a$ and $\Omega.$ \textcolor{black}{Similarly, we present the four classes obtained by lowering the threshold value with respect to the one used in Fig.~\ref{fig:hvsnmin:orient:1tri}. The results are shown in Fig.~\ref{fig:hvsnmin:orient:1tri_4classes}. In panel (b), we display a 2D projection in the plane $V_\theta$, $V_a$ of the four clusters so far obtained. The points have been colored according to their class: class 1 (red) is associated to the ordered states, class 2 (blue) denotes chimera states, while class 3,1 (black) and 3,2 (light gray) correspond to the disordered states. In panel (c), we report the classification as a function of $P$ and $Q$. For each value of the latter, the dominant class is determined from the modal cluster membership of $(V_\theta, V_a, V_\Omega)$. Let us observe that the new class 3,2, offspring of the class 3 in Fig.~\ref{fig:hvsnmin:orient:1tri}, arises only for $P=24$ and sufficiently large $Q$ (light gray points).}

\begin{figure*}[!ht]
\includegraphics[scale=0.145]{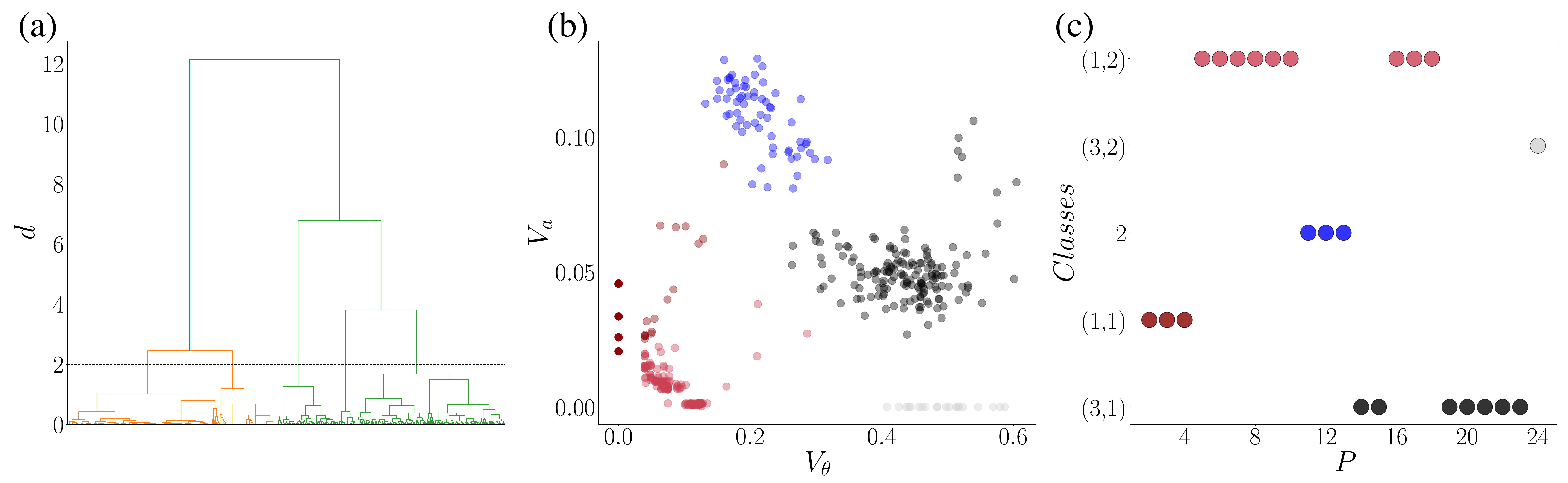}
\caption{\textbf{Even more fine classification of dynamical behaviors by using the variations $V_\theta$, $V_a$ and $V_\Omega$, in the case of orientation $1$ and $Q=P$ links have been reoriented.} In panel (a) we report the dendrogram obtained from the hierarchical clustering obtained by using the values of  $(V_\theta, V_a, V_\Omega)$. Branch lengths represent inter-cluster distances prior to merging, indicating two well-separated groups. With respect to the results shown in Fig.~\ref{fig:hvsnmin:orient:1} we lowered the depth threshold $d$ so to have a classification into five groups. In panel (b) we report the projection in the plane $(V_\theta,V_a)$ of the four obtained clusters: the light red cluster (class 1,1) and dark red cluster (class 1,2) correspond to the ordered states, emerging from the unique regular class shown in Fig.~\ref{fig:hvsnmin:orient:1}, the blue cluster (class 2) corresponds to chimeras, and the black (class 3,1) and light black cluster (class 3,2) correspond to the disordered states. Panel (c) shows an alternative view of the classification as a function of $P=Q$ in the range $\{2,\dots, 24\}$; to each value of $P=Q$ we associated the dominant class, i.e., determined from the statistical modal cluster membership of $(V_\theta, V_a, V_\Omega)$.}
 \label{fig:orient1_5classes}
\end{figure*}

\begin{figure*}[!ht]
\centering
\includegraphics[scale=0.145]{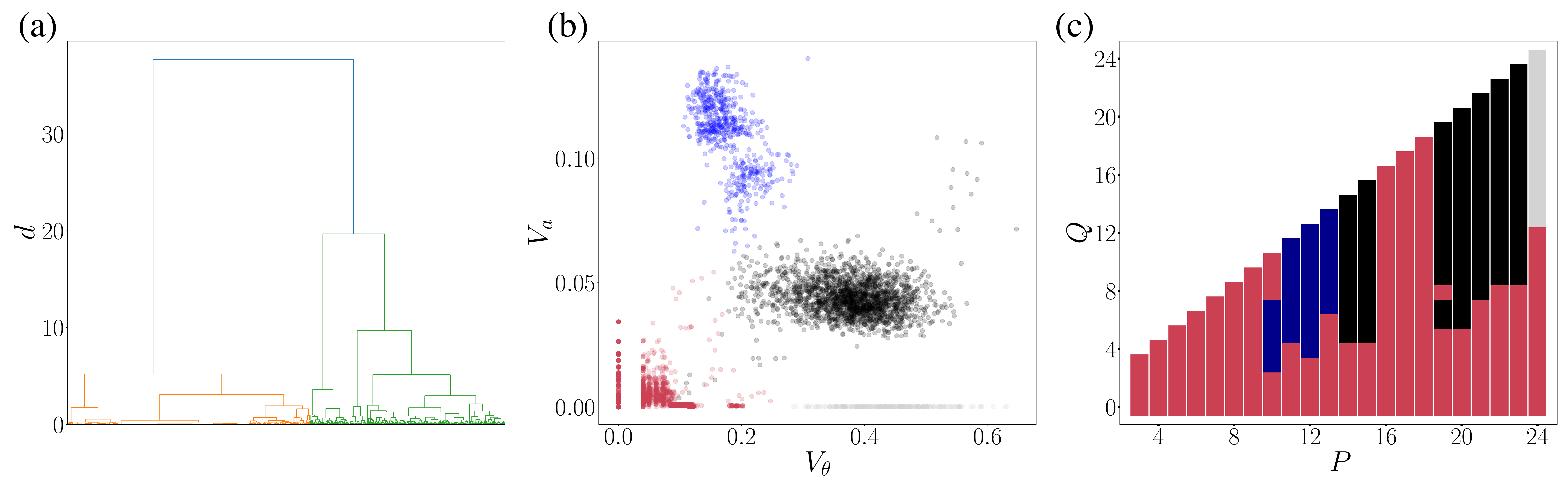}
\caption{\textbf{Finer classification of dynamical behaviors by using the variations $V_\theta$, $V_a$ and $V_\Omega$, in the case of orientation $1$ and $P$, $Q$ links have been reoriented with $0\leq Q\leq P$.} In panel (a) we show the dendrogram obtained from hierarchical clustering of $(V_\theta, V_a, V_\Omega)$; branch lengths represent inter-cluster distances prior to merging, indicating two well-separated groups, we fixed a smaller depth threshold with respect to the one of Fig.~\ref{fig:hvsnmin:orient:1tri} and we thus have four classes. Panel (b) displays a 2D projection in the plane $V_\theta$, $V_a$ of the four clusters so far obtained; points have been colored according to their class: 
 class 1 (red) associated to the ordered states, class 2 (blue) to denote chimera, class 3,1 (black) and 3,2 (light gray) correspond to the disordered states. Let us observe that those two classes were merged in a single class in Fig.~\ref{fig:hvsnmin:orient:1tri}. In panel (c) we report the classification as a function of $P$ and $Q$. For each value of the latter the dominant class is determined from the modal cluster membership of $(V_\theta, V_a, V_\Omega)$. Let us observe that the new class 3,2, offspring of the class 3 in Fig.~\ref{fig:hvsnmin:orient:1tri}, arises only for $P=24$ and sufficiently large $Q$ (light gray points).}
\label{fig:hvsnmin:orient:1tri_4classes}
\end{figure*}

\begin{figure*}[!ht]
\centering
\includegraphics[scale=0.14]{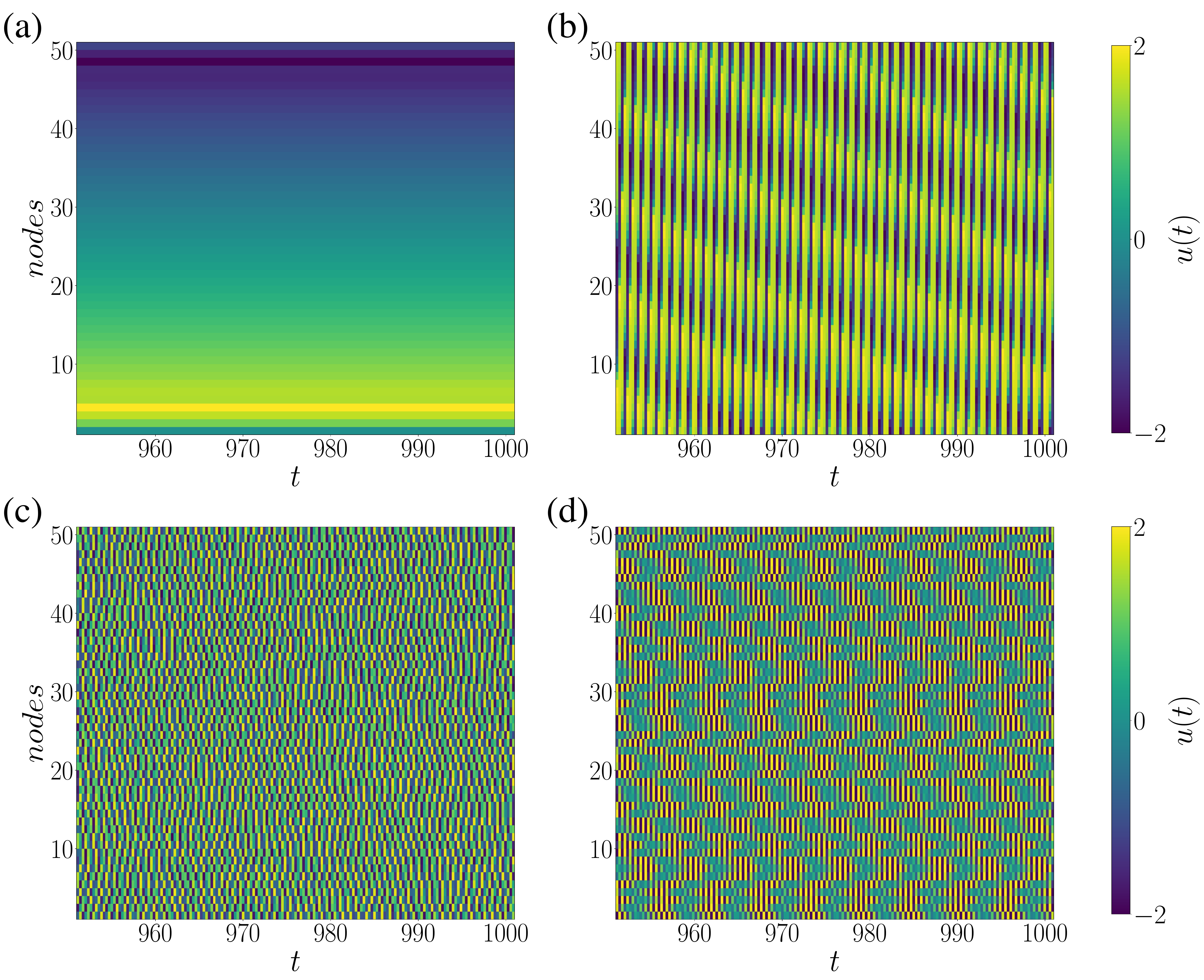}
 \caption{\textbf{Comparison of the dynamics in the case of $5$ classes for orientation $1$}. We show the dynamical behavior of two representative examples in the case of $5$ classes obtained with the finer classification shown in Fig.~\ref{fig:orient1_5classes}. Top panel correspond to the two regular classes, ((a) class (1,1) $P=3$, (b) class (1,2) $P=7$), bottom panels display the irregular classes ((c) class (3,1)  $P=21$, (d) class (3,2) $P=24$). The values of $(V_\theta, V_a, V_\Omega)$ are (0.0000, 0.0336, 0.0000) for (a) showing weakly homogeneous solution , (0.0747, 0.0070, 0.0019) for (b) exhibiting twisted states, (0.4297, 0.0495, 0.0637) for (c)  with irregular behavior and (0.4082,0.00005,0.00002) for (d) showing irregular behavior with very small variation in $a$ and $\Omega.$}
 \label{fig:orient1_otherdyn}
\end{figure*}

\section{Orientation $2$}
\label{sec:orient2}

\begin{figure}[!ht]
\centering
\begin{tikzpicture}[scale=0.80, transform shape]
 \def\radius{3cm}

 \tikzset{
 mnode/.style={
 circle,
 draw=black,
 thick,
 minimum size=0.75cm,
 text=white,
 fill=teal!80, 
 font=\large\bfseries,
 inner sep=0pt,
 outer sep=0.25pt
 },
 midarrow/.style={
 postaction={decorate},
 decoration={
 markings,
 mark=at position 0.6 with {\arrow{Stealth[length=3mm, width=2mm]}}
 }
 }
 }

\begin{scope}[xshift=0cm]
 \foreach \i in {1,...,8} {
 \node[mnode] (n\i) at ({90 - (\i-1)*360/8}:\radius) {\i};
 }

 \foreach \i in {1,...,8} {
 \foreach \k in {1, 2, 3} {
 \pgfmathtruncatemacro{\j}{mod(\i + \k - 1, 8) + 1}
 \pgfmathparse{(\i==8 && \j==1) || (\i== 8 && \j==2) || (\i== 8 && \j==3) || (\i==7 && \j==1) || (\i==7 && \j==2) || (\i==6 && \j==1)}

 \ifnum\pgfmathresult>0
 \draw[thick, red, postaction={decorate, decoration={markings, mark=at position 0.50 with {\arrow{Stealth[reversed]}}}}] (n\i) -- (n\j);
 \else
 \draw[thick, postaction={decorate, decoration={markings, mark=at position 0.50 with {\arrow{Stealth}}}}] (n\i) -- (n\j);
 \fi
 }
 }
\end{scope}

\end{tikzpicture}
\caption{\textbf{Schematic illustration of orientation $2$, for the case $\textcolor{blue}{N_0}=8$ and $P=3$}. This structure has been obtained from the initial ring shown in the left panel of Fig.~\ref{fig:hvsnmin:orient:1} to satisfy the constraint if $\ell=[\textcolor{blue}{i_1},\textcolor{blue}{i_2}]$, then $\textcolor{blue}{i_1}<\textcolor{blue}{i_2}$. More precisely, we fix one node, say $1$, and we reorient links incident to it according to the node index label, i.e., $[1,6]$, $[1,7]$ and $[1,8]$. We then take the node next to $1$, i.e., $2$, and we reorient as many links incident to $2$ as possible, with the constraint about the node index label, in this case $[2,7]$ and $[2,8]$. We then consider node $3$ and we act on its incoming links by reorienting according to the same rule, i.e., $[3,8]$. All links do now satisfy the constraint. The red links indicate those whose orientations differ with respect to the original ring.}
 \label{fig:orient:3}
\end{figure}

{\color{black}
As already stated in the main text, the topology associated to orientation 1 allows to break the rotation symmetry, by changing the orientation of a given number of links, $Q$, but still keeping some regularity. One could be interested in studying the emergence of chimera states once a given number of links is reoriented but in a random way. By anticipating on the following, we will show that chimera states are very sensitive to this disordered reorientation and they will easily disappear working with orientation 1. For this reason we studied a second orientation, named orientation 2, that results to be more robust to induce chimera states, also in the case of random reorientation of links.}

We thus considered a second orientation obtained from the initial ring shown in the left panel of Fig.~\ref{fig:hvsnmin:orient:1} to satisfy the constraint : if $\textcolor{black}{\ell}=[\textcolor{black}{i_1},\textcolor{black}{i_2}]$, then $\textcolor{black}{i_1}<\textcolor{black}{i_2}$ returns a  positively oriented link. A schematic illustration of orientation $2$, for the case $\textcolor{black}{N_0}=8$ and $P=3$ is shown in Fig.~\ref{fig:orient:3}. In this example, we fix one node, say $1$, and we reorient links incident to it according to the node index label, i.e., $[1,6]$, $[1,7]$ and $[1,8]$. We then take the node next to $1$, i.e., $2$, and we reorient as many links incident to $2$ as possible, with the constraint about the node index label, in this case $[2,7]$ and $[2,8]$. We then consider node $3$ and we act on its incoming links by reorienting according to the same rule, i.e., $[3,8]$. All links do now satisfy the constraint. The red links indicate those whose orientations differ with respect to the original ring.

Once the orientation $2$ has been realized, we used it to define the FHN and thus studied its behavior as done for the orientation $1$.
Finally, we classify the dynamics for orientation $2$ as we vary $P$. 
\textcolor{black}{The results are shown in Fig.~\ref{fig:hvsnmin:orient:2_4classes} for depth threshold returning four classes. The latter correspond to: red cluster (class 1) is the ordered states, the blue cluster (class 2) the chimera state, and the clusters (3,1) and (3,2) are associated to the disordered states. We observe chimeras for $P=13,14,15$. Let us notice that here, we have shown a finer classification where the disordered class splits into two disordered ones. If we consider a higher depth threshold as $d=4$, it will return three classes as in Fig.~\ref{fig:hvsnmin:orient:1}, where, i.e., the disordered classes merge into a single one.}


Let us conclude by emphasizing an interesting property of the FHN defined on the ring with orientation $2$ : chimera states are more robust against random reorientation of links as opposed to orientation $1$. To support this claim we considered a ring with orientation $1$ and another with orientation $2$ where in both cases we assume $P=13$. Let us observe that for the choice of the model parameters, those configurations return chimera states (see Fig.~\ref{fig:hvsnmin:orient:1} and Fig.~\ref{fig:hvsnmin:orient:2_4classes}). We then randomly reorient $Q_r$ links, with $Q_r\in\{0,\dots,50\}$, and we determine the presence or lack thereof of chimeras; for each value of $Q_r$ we repeat the construction $T=10$ times, obtaining thus $T$ different random realizations of the orientation $1$ and $2$ with $Q_r$ reoriented links. Let $J$ be the number of times chimera states have been obtained and eventually let $W=J/T$ be the fraction of chimera outcomes for a given $Q_r$. In Fig.~\ref{ratio-o1-o2}, we have plotted the ratio $W$ as a function of the number of reoriented links $Q_r$, teal dots correspond to orientation $2$ and green stars to orientation $1$. We can observe that for small $Q_r$, both orientations return chimera states and indeed $W=1$, however we can appreciate that already at $Q_r = 15 $ orientation $1$ is no longer capable to sustain chimera and indeed $W=0$, on the other hand chimera persist for the orientation $2$ case. Interestingly enough, we can reorient $Q_r=40$ links and still have $W=1$; only for a larger number of reoriented links, chimera states fade out and disappear.

\begin{figure*}[!ht]
\centering
\includegraphics[scale=0.145]{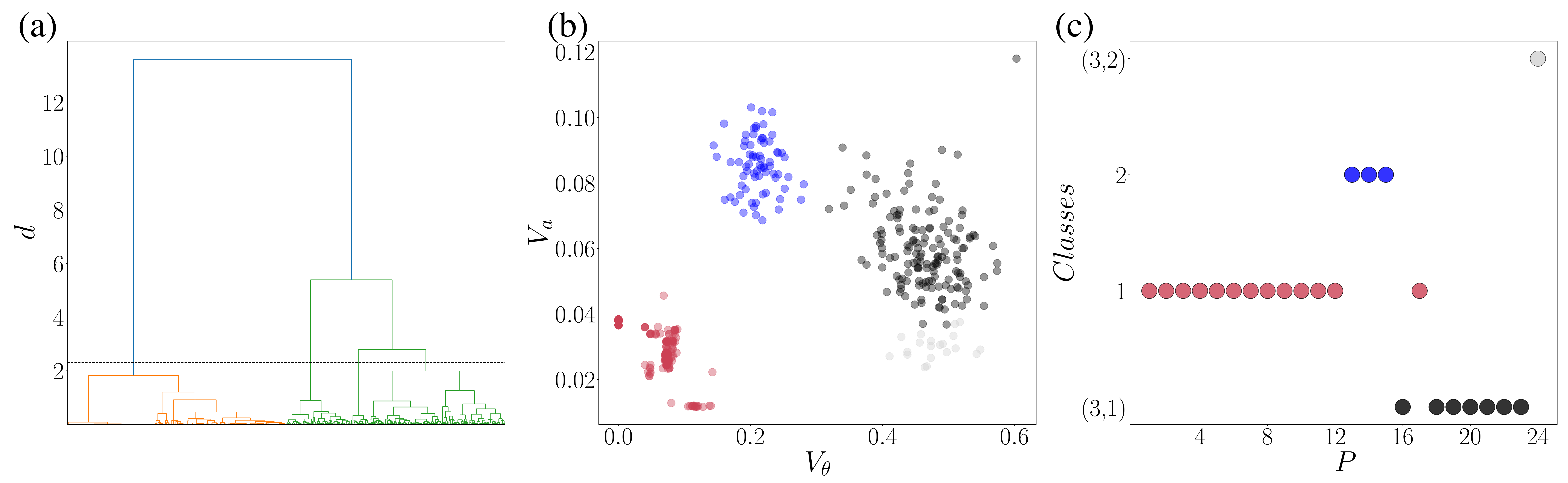}
\caption{\textbf{\textcolor{black}{Classification} of dynamical behaviors by using the variations $V_\theta$, $V_a$ and $V_\Omega$, in the case of orientation $2$.} In panel (a) we report the dendrogram obtained from the hierarchical clustering obtained by using the values of  $(V_\theta, V_a, V_\Omega)$. Branch lengths represent inter-cluster distances prior to merging, indicating two well-separated groups; {\color{black}the threshold depth has been set so to obtain four classes.} In panel (b) we report the projection in the plane $(V_\theta,V_a)$ of the four obtained clusters: the red cluster (class 1) corresponds to the ordered states, the blue cluster (class 2) corresponds to chimeras, and the the black (class 3,1) and light black cluster (class 3,2) correspond to the disordered states, let us observe that they formed a single class in the previous classification. Panel (c) shows an alternative view of the classification as a function of $P$ in the range $\{1,\dots, 24\}$; to each value of $P$ we associated the dominant class, i.e., determined from the statistical modal cluster membership of $(V_\theta, V_a, V_\Omega)$.}
\label{fig:hvsnmin:orient:2_4classes}
\end{figure*}

\begin{figure}
    \centering
\includegraphics[width=0.4\linewidth]{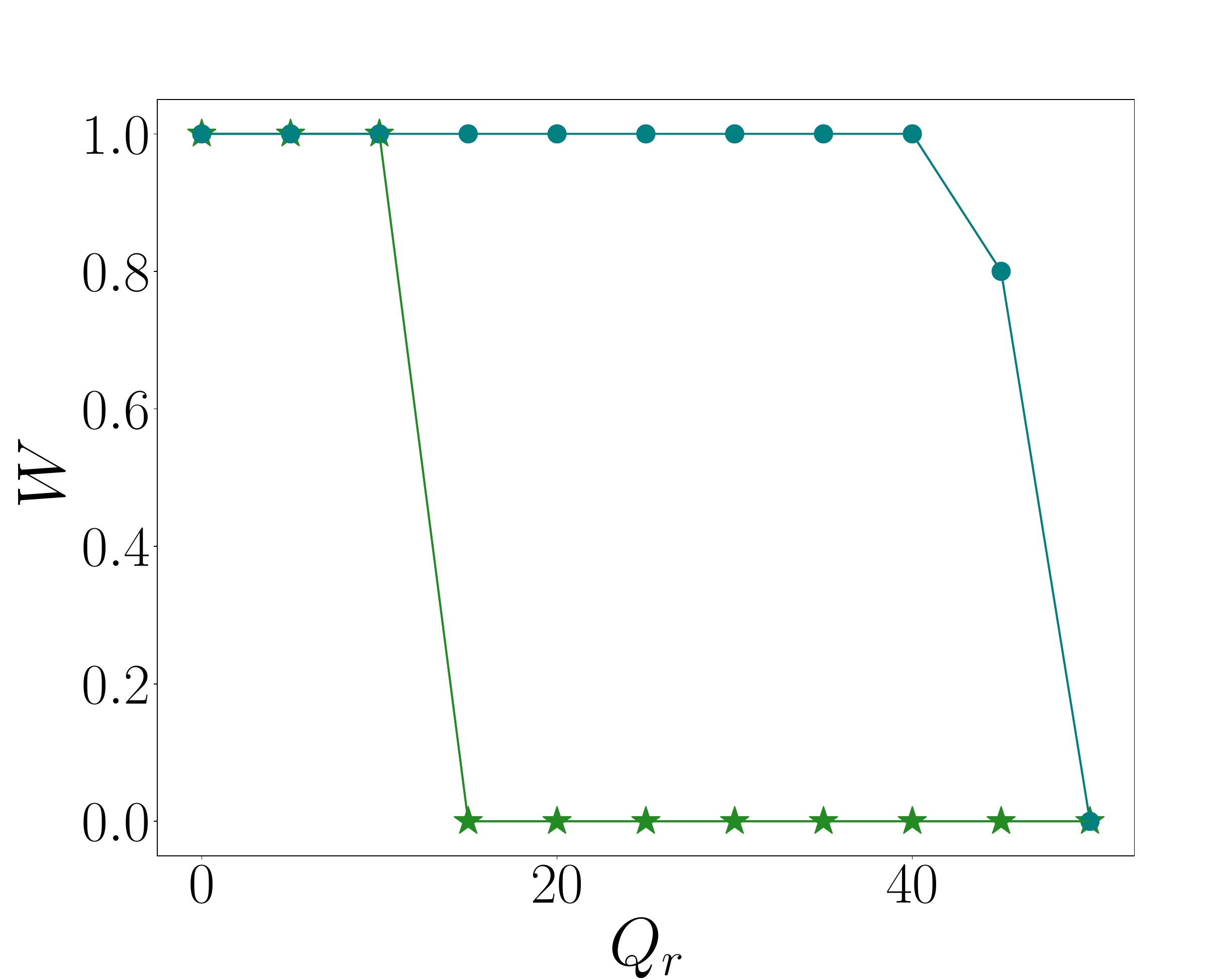}
\caption{\textbf{Robustness of chimera states for orientation $2$ with respect to orientation $1$}. Starting from the orientation $1$ and $2$ cases with $P=13$ where chimeras are present, we reorient $Q_r$ links and we determine the emergence of chimera states. For each $Q_r$ we repeat the process $T$ times and we define $J$ to be number of times chimeras are observed. We thus plot $W=J/T$ as a function of $Q_r$, teal dots correspond to orientation $2$ and green stars correspond to orientation $1$. One can observe that orientation $2$ support chimeras up to $Q_r=40$ while orientation $1$ no longer exhibits chimeras already at $Q_r=15$. We can thus conclude that the chimera states from orientation $2$ are more robust.}
\label{ratio-o1-o2}
\end{figure}

\section{Application to the networked FitzHugh-Nagumo and Comparison with Strength of Incoherence}
\label{sec:compS}

{\color{black}
In the main text we introduced a method to identify chimera states, as well as other dynamical behaviors, in a model of FitzHugh-Nagumo oscillators defined on both nodes and links. As already stated, the method is general enough to be applied to other models; the aim of this section is to apply the classification method to the classical FitzHugh-Nagumo model defined on a network, i.e., where both variables are anchored to nodes and are coupled with the network Laplace matrix instead of the incidence one. Our choice has also been motivated by the sake of comparing the proposed method with the Strength of Incoherence (SI), defined in~\cite{gopal2014} and used to determine the existence of chimera states in the networked FitzHugh-Nagumo model.

The section is thus divided into two parts: first, we introduce the FitzHugh-Nagumo model on a network and we present the results obtained by using our classification method; second, we define the strength of incoherence (SI) and we use it to determine the existence of chimera states in both models. This will allow us to compare the two methods and emphasize their differences.

Let us thus consider a regular ring composed of $N_0$ nodes each one with degree $2P$, being, i.e., connected to $P$ nodes on the right and $P$ on the left. The model equations are thus given by
\begin{eqnarray}
    \label{eq:FHNnet}
    \epsilon\frac{dx_i}{dt}&=&x_i-\frac{x_i^3}{3}-y_i+\frac{\sigma}{2P}\sum_{j=i-P}^{j=i+P}\left[b_{xx}(x_j-x_i)+b_{xy}(y_j-y_i)\right]\notag\\
    \frac{dy_i}{dt}&=&x_i+b+\frac{\sigma}{2P}\sum_{j=i-P}^{j=i+P}\left[b_{yx}(x_j-x_i)+b_{yy}(y_j-y_i)\right]\, ,
\end{eqnarray}
where $x_i$ and $y_i$, for $i=1,\dots,N_0$, are the activator and inhibitor variables, $b$ is the threshold parameter and $\epsilon$ characterizes the time scale. The rotational matrix is given by
\begin{equation*}    \left(\begin{matrix}
b_{xx} & b_{xy}\\
b_{yx} & b_{yy}
\end{matrix}\right)=\left(\begin{matrix}
\cos(\phi) & \sin(\phi)\\
-\sin(\phi) & \cos(\phi)
\end{matrix}\right)\, ,
\end{equation*}
with $\phi\in [-\pi,\pi)$. In the following we fix the parameters to the values $\epsilon=0.05$, $b=0.5$, $\phi=\pi/2-0.1$, $N_0=500$ and $P=rN_0$ with $r=0.33$. Note that those are the values used in the work~\cite{gopal2014}. On the other hand initial conditions have been selected to be uniformly randomly distributed from $U[-1,1]$. 

The system~\eqref{eq:FHNnet} exhibits chimera state for small coupling $\sigma$ and regular behavior for large values of the latter. We thus let $\sigma$ to vary in $[0.1,0.6]$ and for each value we simulate $10$ orbits as above prescribed; we thus eventually apply our classification method, by extracting the dynamical features by using the FFT, compute the normalized total variations and then apply the hierarchical classification method. The results are reported in Fig.~\ref{fig:ourmeth}; panel (a) shows the dendrogram where we emphasize the presence of five classes, for the chosen value of the threshold depth. Panel (b) displays a projection in the plane $(V_{\theta},V_a)$ of the five classes. Finally, panel (c) reports the five classes as a function of $\sigma$, more precisely to each $\sigma$ we associated the most frequent class. We can observe the presence of ordered state (red cluster, class 5), of chimeras of different nature (blue clusters, classes 1,2 and 4), and of the state named multi-chimeras in~\cite{gopal2014} (teal cluster, class 3).
\begin{figure}
    \centering
\includegraphics[scale=0.145]{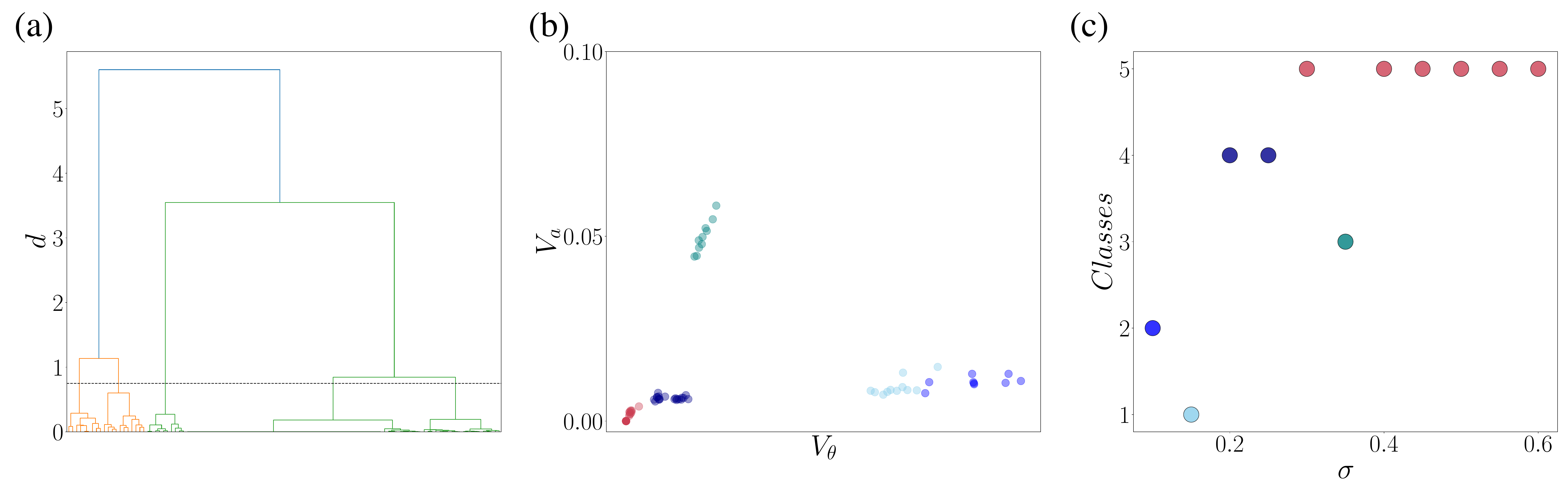}
\caption{{\color{black}\textbf{Classification of dynamical behaviors by using the variations $V_{\theta}$, $V_{a}$ and $V_{\Omega}$, in the case of the network of FHN oscillators}. In panel (a) we report the dendrogram obtained from the hierarchical clustering obtained by using the values of $(V_\theta, V_a, V_\Omega)$. Branch lengths represent inter-cluster distances prior to merging, indicating two well-separated groups; the threshold depth has been set to obtain five classes. In panel (b) we report the projection in the plane $(V_\theta,V_a)$ of the five obtained clusters: the red cluster (class 5) corresponds to the ordered states, the blue clusters (class 1,2 and 4) corresponds to chimeras of different nature, and the teal cluster corresponds (class 3) to a state named multi-chimeras in~\cite{gopal2014}. Panel (c) shows an alternative view of the classification as a function of $\sigma$ in the range $[0.1, 0.6]$; to each value of $\sigma$ we associated the dominant class, i.e., determined from the statistical modal cluster membership of $(V_\theta, V_a, V_\Omega)$.}}
\label{fig:ourmeth}
\end{figure}

Let us now briefly introduce the Strength of Incoherence~\cite{gopal2014}. Starting from the state variable $\mathbf{x}_i=(x_i,y_i)$, we first define a new one, $\mathbf{z}_i = \mathbf{x}_i - \mathbf{x}_{i+1}$, for $i = 1,2,\dots,N_0$. When two neighboring oscillators oscillate coherently, $\mathbf{z}_i$ takes the minimum value, when they oscillate incoherently, it takes the values between, $\pm|x_{i,max} - x_{i,min}|$. To measure the spread of $\mathbf{z}_i$ one can define the long time average of the standard deviation, $\sigma_l$, where $l=1,2$ identifies the component of the state variable we are interested in
\begin{equation}
    \sigma_l = \Big\langle\sqrt{\frac{1}{N_0}\sum_{i=1}^{N_0}[z_{l,i} - \langle z_l\rangle}]^2\Big\rangle_t\, ,
\end{equation} 
One can thus conclude that $\sigma_l = 0$ denotes coherent states and $\sigma_l \ne 0$ for both incoherent and chimera states. Since the latter could not distinguish between the chimeras and incoherent behaviors, authors in~\cite{gopal2014} defined a local standard deviation $\sigma_l(m)$ by dividing the oscillators into an even number $M$ groups of continuous and equal size $n = N_0/M$
\begin{equation}
    \sigma_l(m) = \Big\langle\sqrt{\frac{1}{n}\sum_{j=n(m-1)+1}^{mn}[z_{l,j} - \langle z_l\rangle}]^2\Big\rangle_t\, ,
\end{equation}
$m=1,2,\dots,M$. Eventually, SI is defined as 
\begin{equation}
    S = 1 - \frac{\sum_{m=1}^Ms_m}{M}\, , \, s_m = \Theta(\delta - \sigma_l(m))\, ,
\end{equation}
where $\Theta(.)$ is the Heaviside step function, and $\delta>0$ is a threshold parameter which takes a certain percentage of value of the difference between $x_{l,i,max}$ and $x_{l,i,min}$. In conclusion~\cite{gopal2014} if $\mathrm{SI} = 0$ then the system behavior is regular, $\mathrm{SI} = 1$ for incoherent states and $0<\mathrm{SI} <1$ for chimeras. 

In Fig.~\ref{fig:theirmeth} we report the classification of the dynamical states for the networked FHN model obtained by using the strength of interaction. Panel (a) displays SI computed by using $\delta = 0.002$ and $M=20$, the method correctly identifies regular behavior for $\sigma \geq 0.4$ (red squares) and chimera states for $\sigma <0.4$ (blue squares). Let us however observe that the results strongly depend on the choice of $\delta$, indeed by using $\delta = 0.001$, and again $M=20$, $\mathrm{SI}$ assumes the value $1$ for $\sigma < 0.2$, corresponding thus to irregular behavior, whereas by visual inspection (see panel (c) for the case $\sigma=0.1$) this dynamics is clearly a chimera state. Moreover, for $\sigma=0.3$, SI classifies the state as chimera whereas our method sets it in the regular class (see panel (c) of Fig.~\ref{fig:ourmeth}).
\begin{figure}
    \centering
\includegraphics[scale=0.145]{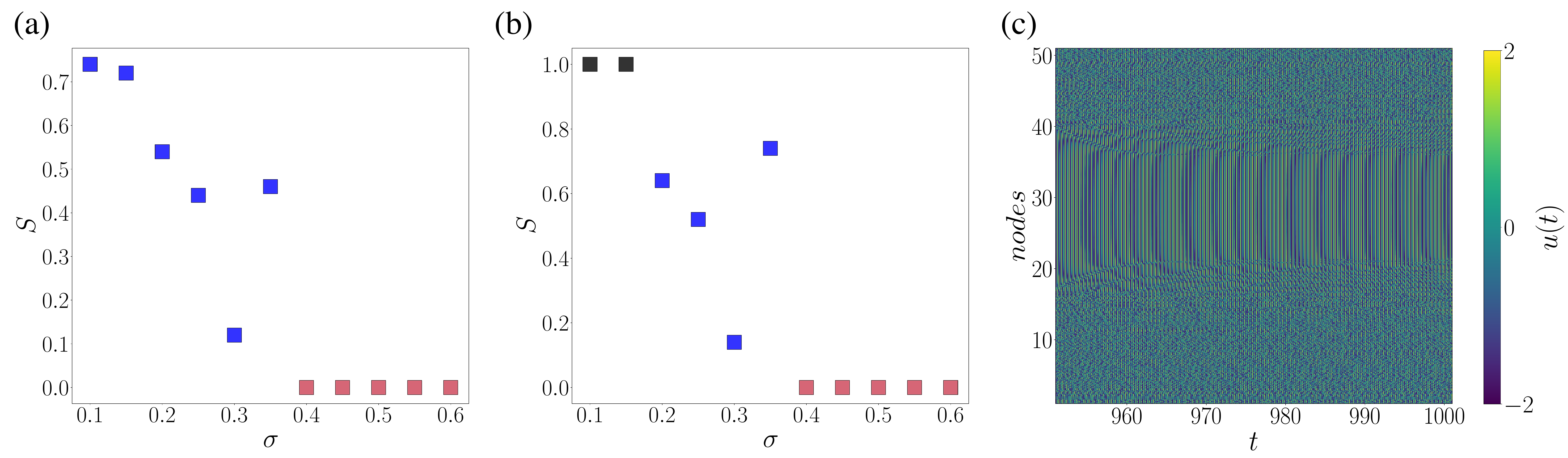}
\caption{{\color{black}\textbf{Strength of interaction (SI) for the network of FHN oscillators}. In panels (a) and (b) we report the SI as a function of $\sigma$, computed by using $\delta = 0.002$ and $\delta = 0.001$ respectively, and $M=20$ in both cases. $\mathrm{SI} = 0$ corresponds to the regular behavior (red squares), $\mathrm{SI} = 1$ corresponds to the incoherent behavior (black squares) and the intermediate values $0<\mathrm{SI}<1$ correspond to chimeras (blue squares). Let us observe that a slight change in $\delta$ causes the misclassifications of the chimeras as irregular behaviors for $\sigma=0.1,0.15$. In panel (c) we present the dynamics associated to $\sigma=0.1$; by visual inspection one can realize it corresponds to a chimera state.}}
\label{fig:theirmeth}
\end{figure}

In Fig.~\ref{fig:Sgoodwrong} we show the results we obtained by using the strength of incoherence to classify dynamical behaviors in the FHN model studied in the main text as a function of $P$, with $Q=P$ for orientation 1. Panel (a) shows SI obtained by using $\delta = 0.1256$ and $M=6$; $\mathrm{SI} = 0$ corresponds to regular behavior (red squares), $\mathrm{SI} = 1$ to incoherent behavior (black squares) and $0<\mathrm{SI} <1$ to chimera states (blue squares). The obtained classification agrees with the one we got by using our method (see panel (c) in Fig.~\ref{fig:hvsnmin:orient:1}), however by slighting changing $\delta$ (see panel (b)) the computed values of SI misclassify some dynamics for $P=2,3,4$ setting them as chimera states instead of regular ones (see panel (c)).

\begin{figure}
    \centering
\includegraphics[scale=0.145]{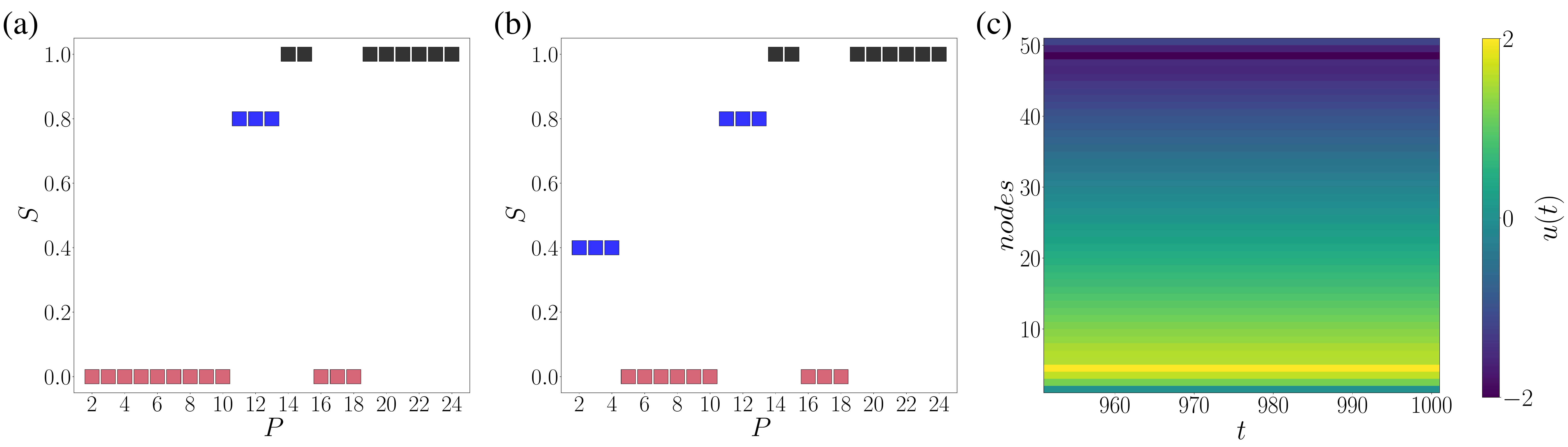}
\caption{{\color{black}\textbf{Classification of dynamical behaviors in the topological FHN model by using the Strength of Incoherence (SI)}. In panels (a) and (b) we report the SI as a function of $P$, with $Q=P$ for orientation 1, for the values $\delta = 0.1256$ and  $\delta = 0.1$ respectively. The value of $M$ is $6$ for the both cases. $\mathrm{SI} = 0$ corresponds to regular behavior (red squares) and $\mathrm{SI} = 1$ corresponds to incoherent behavior (black squares) and the values $0<\mathrm{SI} <1$ correspond to chimeras (blue squares). Again a slight change in $\delta$ causes the misclassifications of the regular behaviors as chimeras for $P=2,3,4$. In panel (c) we present the actual dynamics of $P=Q=3$ which is clearly regular.}}
\label{fig:Sgoodwrong}
\end{figure}

To conclude this section let us consider the metric $\sigma_l$ above introduced and show that it can be inserted into the pipeline of our classification method and produce a good classification scheme without the need to introduce any threshold. This confirms thus the robustness of the proposed scheme and its modulability. We thus consider once again the FHN model defined in the main text for orientation 1, we vary $P$, with $Q=P$; we numerically compute $20$ orbits with different initial conditions and for each one we computed $\sigma_l$, with $l=1$. Because we now use a single variable, we can use an histogram to classify data. More precisely we use the Kernel Density Estimation (KDE)\cite{wkeglarczyk2018kernel} method to determine in a self-consistent way the number of classes that results to be three: regular, chimera and irregular behavior (see panel (a) of Fig.~\ref{fig:sigmal}); let us observe that the obtained classification (see panel (b) of Fig.~\ref{fig:sigmal}) agrees with the one presented in the main text (see panel (c) of Fig.~\ref{fig:hvsnmin:orient:1}).
\begin{figure}
    \centering
\includegraphics[width=0.7\linewidth]{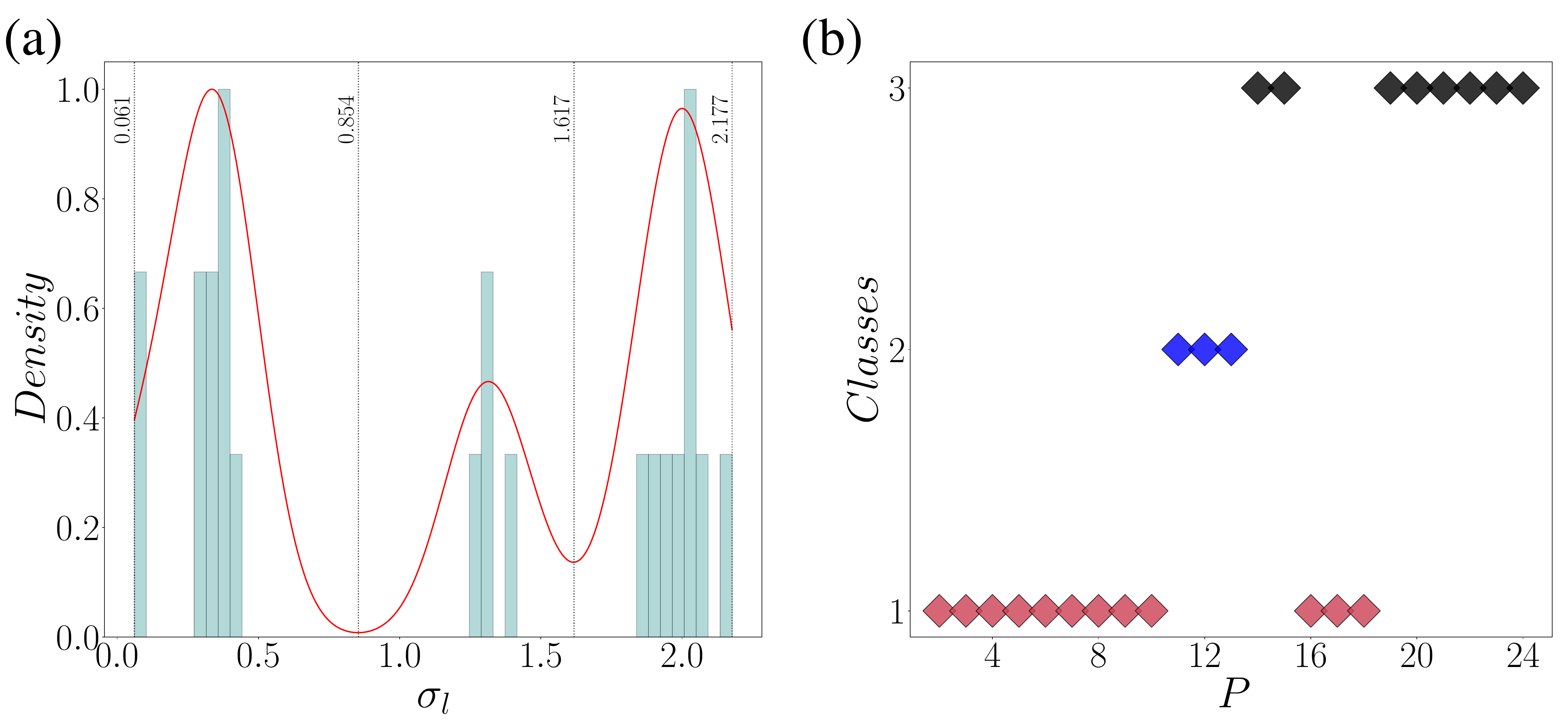}
\caption{{\color{black}\textbf{Classification of dynamical behaviors in the topological FHN model by using $\sigma_l$}. In panel (a) we present the histogram of the computed values of $\sigma_l$, $20$ for each value of $P$, and the three classes obtained by using Kernel Density Estimation (red curve), each class being identified by the vertical lines. In panel (b) we report the classes obtained by using $\sigma_l$ as a function of $P$. The red diamonds correspond to the regular behavior, the blue diamonds are chimeras and the black diamonds correspond to the incoherent behavior. The results agree with the classification obtained by using $(V_\theta,V_a,V_\Omega)$ as shown in Fig.~\ref{fig:hvsnmin:orient:1}.}}
\label{fig:sigmal}
\end{figure}
}
\end{document}